# Laboratory investigations of the bending rheology of floating saline ice, and physical mechanisms of wave damping, in the HSVA ice tank

Aleksey Marchenko[1], Andrea Haase[2], Atle Jensen[3], Benjamin Lishman[4], Jean Rabault[3], Karl-Ulrich Evers[5], Mark Shortt[6], Torsten Thiel[7]

[1]The University Centre in Svalbard, Norway

[2]Arctic Technology, Hamburgische Schiffbau-Versuchsanstalt GmbH, Hamburg Ship Model Basin, Germany

[3]University of Oslo, Norway

[4]London South Bank University, UK

[5]Solutions4arctic, Germany

[6]University College London, UK

[7]Advanced Optics Solutions GmbH, Germany

## Abstract

An experiment on the propagation of flexural-gravity waves was performed in the HSVA ice tank. Physical characteristics of the water-ice system were measured in different locations in the tank during the tests, with a number of sensors deployed in the water, on the ice and in the air. Water velocity was measured with an acoustic doppler velocimeter (ADV) and an acoustic doppler current profiler (ADCP); wave amplitudes were measured with ultrasonic sensors and the optical system Qualisys; in-plane deformations of the ice and the temperature of the ice and water were measured by fiber optic sensors, and acoustic emissions were recorded with compressional crystal sensors. All together 61 tests were performed, with ice thicknesses of 3 cm and 5 cm. The experimental setup and selected results of the tests are discussed in this paper. We show that cyclic motion of the ice along the tank, imitating ice drift, causes an increase in wave damping. We also show that the formation of non-through cracks in the ice, caused by the action of waves, increases wave damping.

## 1. Introduction

Sea ice coverage in the Arctic Ocean is decreasing. This leads to an increase in the probability of storms in ice-free areas (Sepp and Jaagus, 2011). Surface waves and swell penetrate from stormy regions of the ocean into ice-covered regions, and induce ice failure. Between open water and solid ice, the waves pass through the marginal ice zone (MIZ), which consists of broken ice. The marginal ice zone acts as a low pass filter to the waves, with characteristics which depend on floe sizes and the concentration of ice on the water surface (Wadhams, 2000). The nature and extent of the MIZ determine the characteristics of the waves which reach regions covered by solid ice. Since the ice failure is influenced by the waves, and the failed ice controls the ways the waves are filtered, the ice and waves are linked in a feedback mechanism. As well as waves, the structure of the MIZ is influenced by other factors, including wind, surface current and sea surface tilt. Waves can influence the characteristics of the MIZ over large areas in a relatively short time, i.e. several hours (see e.g. Collins et al., 2015). Safe operations in the Arctic require high-precision forecasts of sea ice. The description of wave-ice interactions in models of weather and sea-state forecasts is therefore important to all maritime endeavours in the Arctic.

In-situ observations of wave-ice interactions in the MIZ have been widely described (see e.g. Robin, 1963; LeSchack and Haubrich, 1964; Wadhams et al, 1986; Martin and Becker, 1987; Wadhams et al., 1988; Smirnov, 1996; Doble and Wadhams, 2006; Hayes and Jenkins, 2007; Kohaut et al., 2015; Marchenko et al., 2017; Tsarau et al, 2017a). In most of these field observations, waves were recorded in the MIZ with periods from 5s to 20s, i.e. in the spectral range of wind waves and swell. Floe thicknesses range vary within the MIZ and can reach 3-4m for individual floes. Ice concentration increases from the zone edge to the region covered by solid ice. The attenuation of wave amplitudes in the MIZ is exponential with distance into the ice, and the coefficient of attenuation in the exponent increases with decreasing wave period. There can be a roll over effect when the natural frequencies of floe oscillations are excited by incoming waves (Wadhams et al., 1988). Physical mechanisms of wave damping in the MIZ include wave scattering at floe edges (see e.g. Squire et al., 1995), nonelastic floe-floe interactions, and energy dissipation in the under-ice boundary layer. Characteristics of wave-induced floe-floe collisions are discussed by e.g. Martin and Becker (1987), Shen and Squire (1998), Frankenstein et al. (2001), and Doble and Wadhams (2006). Weber (1987) considered wave damping due to energy dissipation in the under-ice boundary layer. His estimates, derived from the theory of laminar boundary layers near an oscillating plate (Lamb, 1932), predict very little damping. Liu and Mollo-Christensen (1988) developed these ideas, suggesting that eddy velocity should be used (rather than molecular velocity) to describe the oscillating boundary layer. Kohaut et al. (2011) derive a relation for wave attenuation due to drag in steady flow (building on observations from Langleben, 1982). Marchenko et al. (2015) analyzed several wave propagation events below drift ice in the Barents Sea, and estimated swell damping using the eddy viscosity calculated from in situ measurements of water velocity fluctuations in the ice-adjacent boundary layer. High values of eddy viscosity – above 100 $cm^2s^{-1}$ – were found in cases when the ice drifted with relatively high speeds. The influence of waves on eddy velocity was not discovered. As a side note this also agrees with measurements performed in grease ice close to the shore, where intense eddy activity may also be present (Rabault et al., 2017). Such eddy structures were later measured in small scale experiments in the laboratory (Rabault et al., 2018).

The low frequency component of swell propagates across long distances under the ice without ice failure and with very little damping, causing bending oscillations of Arctic pack ice. Measurements of swell in Arctic pack ice have been made in the Beaufort Sea (Crary et al., 1952) and in the Central Arctic (Hunkins, 1962; LeSchack and Haubrich, 1964; Sytinskii and Tripol'nikov, 1964), using gravity-meters and seismometers. Recently, Mahoney et al., (2016) measured low frequency swell using short-temporal-baseline interferometric synthetic aperture radar. The results of these various swell measurements (all made in Arctic pack ice) are summarized in Table 1. Hunkins (1962), Sytinskii and Tripol'nikov (1964), Gudkovich and Sytinskii (1965) and Smirnov (1996) measured waves with periods of 8-15s. These are associated with local processes in drift ice, caused by wind action on ice ridges, floe-floe interactions, etc. Physical mechanisms of wave damping in solid ice include viscous and anelastic bending deformations of ice, energy dissipation in the ice-adjacent boundary layer, and brine pumping (Marchenko and Cole, 2017).

Table 1. Characteristics of low frequency swell in the Arctic Ocean

| | Sea Depth, km | Ice thickness, m | Wave amplitude, mm | Wave period, s |
|---|---|---|---|---|
| Crary et al, 1952 | 3.4-3.8 | - | 0.5 | 5-40 |
| Hunkins, 1962 | >1 | 3 | 5 | 15-60 |
| LeShack and Haubrich, 1964 | 3 | 1-3 | 0.5 | 20-60 |
| Sytinskii and Tripol'nikov, 1964 | >1 | 3 | 0.5 | 20-40 |
| Mahoney et al., 2016 | 0.15 | - | 1.2-1.8 | 30-50 |

Wave actions on pack ice and land-fast ice are similar because both involve solid ice. It is relatively easy to organize and conduct field work on land-fast ice since it is connected to the shore line. The action of the land-fast ice on incoming waves is combined with the effects of bathymetry, shoreline and islands. These combined effects can lead to diffraction, refraction and reflection of waves, leading to waves with more complicated configurations. The amplitude of ocean swell in shallow water regions becomes higher, and the wavelength becomes shorter. This swell can also cause the breakup of land-fast ice near the shoreline. Zubov (1944) describes the breakup of landfast ice near Cape Chelyuskin and Tiksi Bay on 26th-28th January 1943 by large waves, despite the fact that the ice thickness throughout the Laptev Sea was greater than 1m. Bates and Shapiro (1980) recorded vertical displacements of several centimeters amplitude, and with a period around 600s, prior to a significant ice push episode in land-fast sea ice (1.5-2m thick) near Point Barrow, Alaska. Further, over five years of near-continuous radar observations of near-shore ice motion in that area, similar oscillations were always observed to occur for several hours before the start of movement of land-fast ice or adjacent pack ice. Wave events were associated with momentum transfer from sea ice into the water during ice ridge buildup between land-fast ice and drift ice in the Beaufort Sea (Marchenko et al., 2002). The breakup of land-fast ice of 0.5m thickness, in shallow water near the shore, due to swell with period 7s and amplitude 10-15cm, is described by Marchenko et al. (2011). This work also shows that the maximum bending stresses in the ice during the breakup event were comparable to the flexural strength measured in the same place several days earlier. The action of a tsunami-wave on land fast ice (1m thick, near Tunabreen glacier in Temple Fjord, Spitsbergen) is described by Marchenko et al. (2012, 2013). The duration of the leading wave pulse was 40s, and the wave tail included waves with periods around 10s and 16s. Sutherland and Rabault (2015) investigated how swell penetrates from open water into land-fast ice (0.5m thick, in Temple Fjord, Spitsbergen), and were able to measure the attenuation of waves, with periods 4-10s, in the land-fast ice.

To model wave-ice interactions we need a mechanical model of bending deformation. Sea ice rheology is well-investigated, in particular for uniaxial loading at constant temperature (see, e.g., Schulson and Duval, 2009). This rheology is characterised by elastic properties, creep properties and delayed elasticity (anelastic properties). Elastic moduli for ice Ih are in the range 3-14 GPa, and Poisson's ratios are in the range 0.274-0.415 at -16°C (Gammon et al., 1983). In many applications, sea ice is considered as an isotropic material with elastic properties characterised by the effective elastic modulus and Poisson's ratio. The effective elastic modulus depends on temperature, salinity and gas content. It also depends on the method of measurement (see e.g. Weeks, 2010). Values of effective elastic moduli measured by static and dynamic methods are in the range of 1-10 GPa. In-situ experiments on the bending of floating cantilever beams in Spitsbergen fjords and in the Barents Sea show that the effective sea ice elastic modulus averaged over the ice thickness is within the range of 1-2 GPa (Marchenko et al, 2017). For Poisson's ratio, a representative value of 0.33 is often used. Timco and Weeks (2010) write that "the effective Poisson's ratio for sea ice is still very poorly understood. There are a large number of factors that influence its value including the loading rate, temperature, grain size, grain structure, loading direction, state of microcracking, etc."

Wadhams (1973) used Glen's model (1955) to estimate the influence of creep on wave damping during wave propagation under solid ice. The observed attenuation rates of waves in ice are fitted best by a Glen-type flow law with an exponent n= 3 and creep parameter similar to the laboratory value for polycrystalline ice. Glen's model describes secondary creep of ice which develops on time scales much larger than the wave periods. In the absence of dynamic stress experiments on ice, an analogy with metal behaviour was used to provide physical context for the model. Wadhams (1973) noted that experimental confirmation of the model, in cyclic creep experiments on ice, would be useful. Squire and Allan (1980) used a linear viscous-elastic model to describe bending deformations of floating ice. Using the variational approach of Biot (1955) they derived a general equation for the flexural bending of an isotropic linear viscoelastic thin plate floating on the surface of an ideal fluid. Numerical values for the rheological constants were taken from Tabata (1958).

Most experiments on creep and anelasticity have been performed for relatively long-term loading, with representative time much greater than the wave periods (e.g. Budd and Jacka, 1989). Experiments by Cole (1995), performed with a cyclic frequency of 1Hz and lower, show that elastic rheology dominates when the load amplitudes are small enough and dislocations are not growing in the ice. Creep and anelastic properties of ice cause a phase shift between strains and stresses. Further, the specific bending rheology of floating ice is related to the vertical temperature and salinity gradients in the ice: the temperature at the bottom of the ice is equal to the freezing point, and the temperature at the top of the ice is lower. Numerical estimates with Cole's model show that the amount of dissipated energy is not greater than 5% of the elastic energy of ice subjected to bending deformations (with maximum stresses below 0.5MPa) (Marchenko and Cole, 2017).

Wave-induced ice break-up criteria are based on models of flexural strength and fatigue. Flexural strength determines the conditions under which the ice is broken by static bending. The flexural strength of ice depends on temperature, salinity and gas content. Timco and O'Brien, 1994, summarize results of about 2500 laboratory and in situ tests of sea ice flexural strength. The fatigue characteristics of sea ice give an indication of its possible failure mechanisms under repeated loading. Unique full-scale cyclic loading tests, performed in Antarctica on floating cantilever beams of 2m thick ice, are presented by Haskell et al. (1996). In addition, the results of more than 60 full scale tests with floating cantilever beams of sea ice, as well as laboratory and in-situ small scale tests, performed in Spitsbergen Fjord and in the Barents Sea, are summarized by Marchenko et al. (2017).

Laboratory tests, with ice made in the laboratory, are widely used to investigate ship-ice interaction and ice actions on structures (see e.g. Ashton, 1986). To interpret the results of these tests, it is necessary to formulate and use scaling laws for model tests with waves in ice. Laboratory tests can be performed with different model materials, which imitate ice and may satisfy certain similarity criteria. Ice-ship and ice-structure interactions are modelled by ice tank laboratory tests in spite of the high cost of these tests (see e.g. Evers, 2017). Model ice is a material choice which reproduces many observed scenarios and behaviours in interactions, ice strength criteria, and characteristics of ice-structure friction. Wave diffraction by floes, and the drift of floes, can be investigated with model floes made from other materials with appropriate buoyancy and elasticity (see., e.g., Ofuya and Reynolds, 1967; Sakai and Hanai, 2002; Kohout et al., 2007; Prabovo et al., 2014). However, only a few laboratory studies of wave propagation under ice sheets have been conducted (Evers and Reimer, 2015). Squire (1984) describes experiments on wave penetration below the ice in a laboratory flume (2m long, 1m wide, and 0.6m in depth), using natural polycrystalline ice with thickness 3-4cm, and wave periods from 0.6-0.8s. These experiments showed that the amplitude of the vertical acceleration of the ice decreases with distance from the ice edge.

In 2015 and 2016, several tests on wave-ice interaction were performed at the Large Ice Model Basin (LIMB) of the Hamburg Ship Model Basin (Hamburgische Schiffbau-Versuchsanstalt, or HSVA) (Cheng et al, 2017; Tsarau et al., 2017b; Hermans et al., 2018). The main goals of these tests were (1) to investigate the distribution of floe sizes when an initially continuous uniform ice sheet was broken by regular waves with prescribed characteristics, (2) to measure wave attenuation and dispersion in broken ice, and (3) to improve understanding of ice-structure interaction under wave conditions. Wave characteristics were reconstructed from the records of water pressure sensors mounted on the tank wall. Tests were performed with wave lengths around 2.5m and 6.17m. Both ice breakup (starting from the ice edge) and wave attenuation were observed in the tests with wave length around 2.5m. The width of the broken region reached 22m but did not extend over the entire ice sheet.

In this paper we present the experimental setup and broad results of tests performed in January 2018 in the Large Ice Model Basin (LIMB) of HSVA. The work was supported by the Hydralab+ project "Investigation of bending rheology of floating saline ice and physical mechanisms of wave damping". The aims are: to observe and describe physical processes in ice during wave propagation; to investigate the bending rheology and failure conditions of floating solid ice; and to investigate the damping of waves propagating below solid and fractured ice. These aims are realized by performing a suite of measurements during wave propagation below ice. The measurements include: elevation of the ice surface; water pressure under the ice; in-plane strains in the ice; point and profile measurements of water velocity below the ice; acoustic emissions from the ice; ice and water temperature; and properties of the waves in open water. The paper is structured as follows. Section 2 provides a description of the experimental setup. Section 3 discusses similarity criteria and scaling. Section 4 discusses visual observations of mechanisms of wave damping and ice failure in the ice tank. Specific descriptions of sensors and their deployment, along with examples of recorded data, are given in sections 5-10. Section 11 gives a discussion of these test results and outlines plans for future investigation.

## 2. Experimental setup

The experimental programme was focused on the investigation of surface wave propagation below solid ice. Therefore, 38 tests were performed with solid ice, 1 test with the ice split into square blocks manually, and 2 tests with the ice broken by waves. Two groups of tests (TGI and TGII) were conducted during the test programme. The ice thickness was 3 cm in TGI and 5 cm in TGII. Measurements were performed with the sensors listed in Table 1. Locations of the installed sensors are shown in Fig. 1. A Qualisys™ motion capture system is used to detect the rigid body motions of the ice in all six degrees of freedom (6-DOF). The system uses four cameras, installed on the main carriage, to detect markers which are located at different positions on the model. The locations of AE, FBGS and FBGT sensors were different in TGI and TGII because of technical constraints. The locations of the other sensors were unchanged.

Table 1. List of sensors, their short names and symbols.

| Sensors (Full names) | Abbreviated name | No. of sensors | Symbol in Fig.1 |
|---|---|---|---|
| Acoustic Doppler Current Profiler | ADCP | 3 | □ |
| Acoustic Doppler Velocimeter | ADV | 2 | □ |
| Acoustic Emission Transducers | AE | 8 | ■ |
| Fiber Bragg Grating Strain Sensors | FBGS | 8 | ▯ |
| Fiber Bragg Grating Thermistor String | FBRGT | 2 | ▯ |
| Qualisys™ | Q | 6 | ● |
| Ultrasonic sensors | US | 24 | ▬ |
| Water pressure sensor | WP | 8 | ▯ |

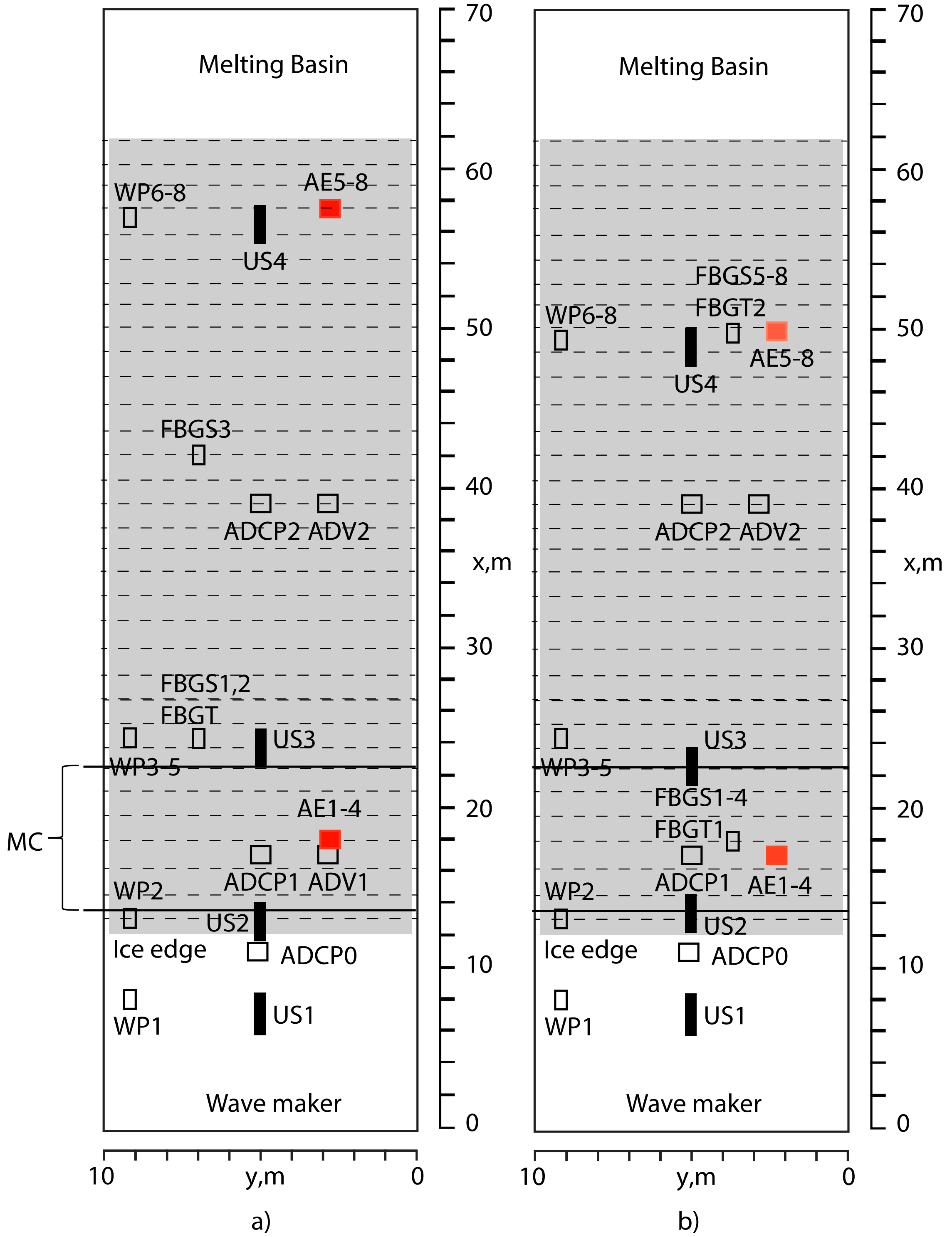

Figure 1. Locations of the sensors in TGI (a) and TGII (b). Designations are given in Table 1.

TGI and TGII included similar tests performed with varying wave frequencies in the range from 0.7 Hz to 1 Hz and varying open water wave heights in the range from 0.5 cm to 1.5 cm. These tests were repeated on (a) steady ice and (b) ice cyclically moving along the *x*-direction with an amplitude of about 1 m. Cyclic motion of the ice was produced manually by two persons pushing and pulling the entire ice sheet along the tank using two poles with hooks. TGI finished with tests performed on manually broken ice with rectangular floes (1.8x1.8 m). TGII finished with tests on the ice after it had been broken by waves.

The model ice cover has salinity of 2.8-3.2ppt and consists of two layers. During the experiment campaign from 15 January to 18 January 2018 the salt content in the ice decreased from about 3.2ppt to 1.6ppt due to drainage of brine. The upper layer, of about 5 mm thickness, consisted of granular crystals as a result of the seeding process. The average grain diameter is about 1 mm. Thereafter, the ice continues to grow, forming relatively long columnar crystals (Fig. 2). These crystals reach a diameter of about 2-4 mm at the bottom of the 50 mm thick ice sheet. Air is pumped into the water during the ice growth, such that micro air bubbles of 200-500μm diameter are trapped by the ice crystals and are distributed homogeneously in the ice cover.

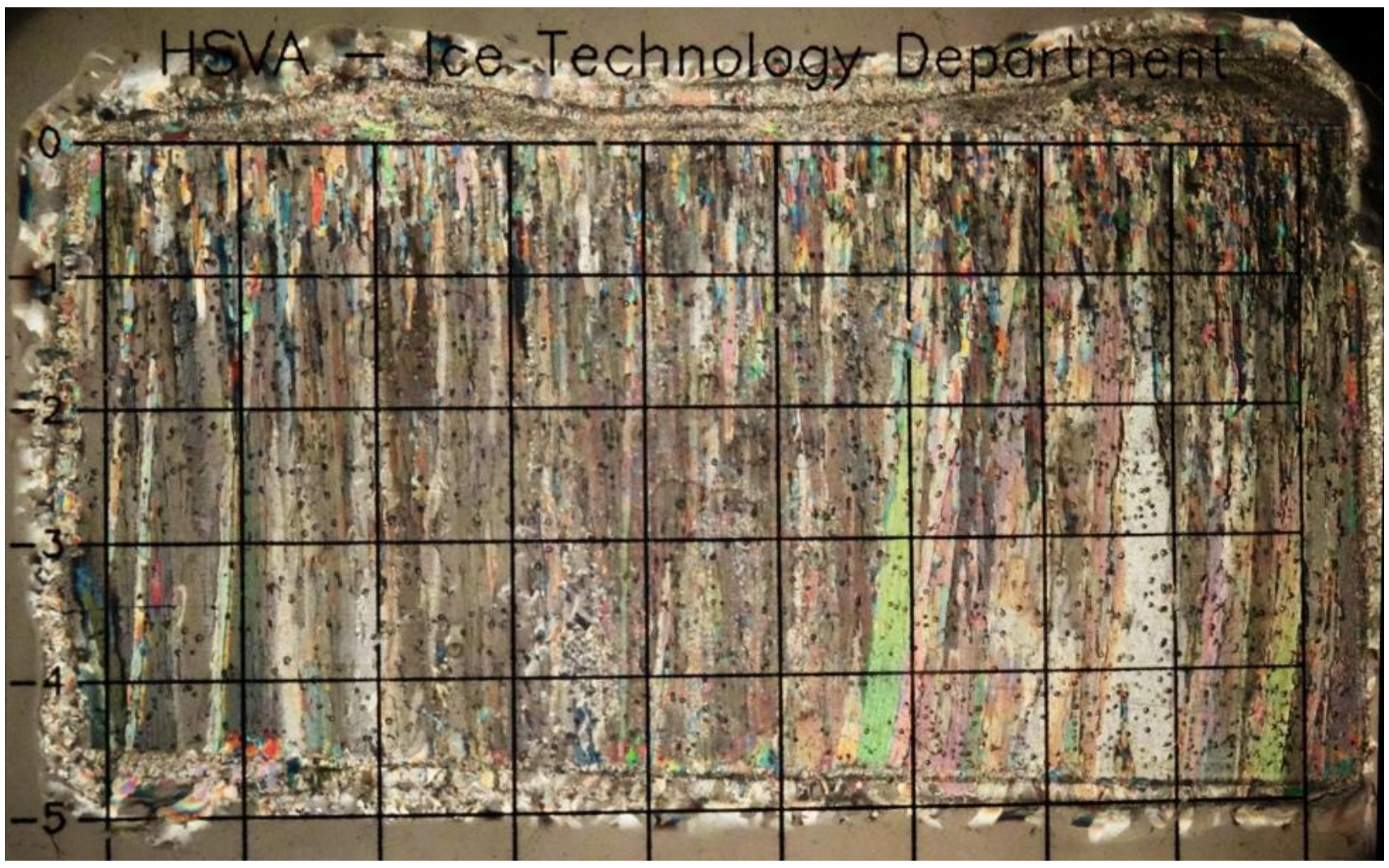

Figure 2. Thin section of model ice.

The elastic modulus and flexural strength of model ice was measured each day using a point loading method and flexural strength tests with floating cantilever beams. The mean values of the elastic modulus and flexural strength measured in TGI and TGII are shown in Table 2. 'AT' and 'BT' denote values measured before and after the tests. There is a reduction of flexural strength during TGII.

Table 2. Elastic moduli and flexural strengths measured during TGI and TGII

| TGI | | TGII | | | |
|---|---|---|---|---|---|
| *E*, MPa | $\sigma_f$, kPa | *E*, MPa (BT) | *E*, MPa (AT) | $\sigma_f$, kPa (BT) | $\sigma_f$, kPa (AT) |
| 43 | 92 | 250 | 245 | 118 | 86 |

The schedule of tests is shown in table 3. The wave heights given in the table correspond to the waves produced by the wave maker in open water. These wave heights are programmed before the wave maker starts to work. The ice sheet affects actual wave heights in the tank due to damping in the ice covered region (8 m < $x$ < 62 m) and due to reflections from the ice edge in the region with open water (0 < $x$ < 8 m). Wave reflection from the end of the tank can be ignored in the tests because of the small wave amplitudes at the end of the tank and the relatively small wave lengths.

Table 3. Test matrix of selected tests.

| Dates | Names | Frequency, Hz | Wave height, mm | Comments |
|---|---|---|---|---|
| 16.01;14:30 | T16_05 | 0.5 | 10 | Fixed ice |
| 16.01;14:51 | T16_06 | 0.6 | 10 | Fixed ice |
| 16.01;15:32 | T16_08 | 0.8 | 10 | Fixed ice |
| 16.01;16:12 | T16_10 | 1.0 | 10 | Fixed ice |
| 16.01;17:42 | T16_06_mov | 0.6 | 10 | Moving ice |
| 16.01;18:02 | T16_08_mov | 0.8 | 10 | Moving ice |
| 16.01;18:22 | T16_10_mov | 1.0 | 10 | Moving ice |
| 17.01;11:28 | T17_10 | 1.0 | 10 | Fixed ice |
| 17.01;11:50 | T17_08 | 0.8 | 10 | Fixed ice |
| 17.01;12:11 | T17_06 | 0.6 | 10 | Fixed ice |
| 17.01;14:08 | T17_10_mov | 1.0 | 10 | Moving ice |
| 17.01;14:30 | T17_08_mov | 0.8 | 10 | Moving ice |
| 17.01;14:50 | T17_06_mov | 0.6 | 10 | Moving ice |
| 17.01;16:19 | T17_07 | 0.7 | 10 | Fixed ice, non-through crack is formed |
| 17.01;16:40 | T17_07_01 | 0.7 | 10 | Fixed ice, non-through crack is formed |
| 17.01;18:02 | T17_07_02 | 0.7 | 20 | Fixed ice, non-through crack develops into through crack |
| 18.01;10:28 | T18_07 | 0.7 | 15 | Fixed ice; the ice is broken in the region x<30 m |
| 18.01;10:57 | T18_07_01 | 0.7 | 30 | Fixed ice; the ice is broken in the region |
| | | | | x<30 m |
| 18.01;11:28 | T18_10 | 1.0 | 30 | Fixed ice; the ice is broken in the region x<30 m |
| 18.01;14:30 | T18_07_02 | 0.7 | 60 | Fixed ice; ice continues to break further downstream |

3. Similarity criteria and scaling

In order to interpret the test results, it is useful to formulate scaling laws for model tests with waves in ice. Specifically, this helps to clarify which naturally-occurring wave-ice interactions are comparable to those in the tests described in this paper. In addition to the Froude ($Fr = V/\sqrt{gh}$) and Cauchy ($Ch = \rho_w V^2 / E$) numbers, and wave slope ($ak$) used for the scaling of water-ship interaction, a set of dimensionless parameters includes

$$\rho_i / \rho_w,\ \nu V / gh^2,\ \sigma_f / \rho_w gh,\ E / \sigma_f,\ \kappa V / \nu h, \tag{1}$$

where $\rho_i$ and $\rho_w$ are the water and the ice densities, $a$ and $k$ are the amplitude and wave number, $\nu$ is the kinematic viscosity of water below the ice, $V$ is either phase either group wave velocity, $g$ is the acceleration due to gravity, $h$ is the ice thickness, $E$ is the effective elastic modulus of ice, $\sigma_f$ is ice flexural strength, and $\kappa$ is the permeability of ice. For broken ice, the ratio $l_f / h$, where $l_f$ is a representative diameter of floes, should be added as geometrical scaling parameter. Another geometrical parameter, $h/H$, is added when the water depth $H$ influences wave properties. Additional parameters proportional to rheological constants should be added to set (1) when the influence of viscous and anelastic properties of ice is important. In the experiments where the ice doesn't fail, the use of the dimensionless parameters which include flexural strength is not necessary.

The dispersion equation describing flexural-gravity waves, ignoring ice inertia, has the form

$$\omega^2 = gk \tanh(kH)\left(1 + Dk^4\right),\ D = \frac{Eh^3}{12(1-\nu^2)\rho_w g}, \tag{2}$$

where $\nu$ is Poisson's ratio, which for sea ice is typically between 0.3 and 0.4 (see, e.g., Timco and Weeks, 2010). For the estimates, here it is assumed that $1-\nu^2 \approx 1$. Equation (2) gives the dispersion equation of gravity waves when $D = 0$. Dispersion curves of flexural-gravity waves (FGW1 and FGW2) and gravity waves (GW) are shown in Fig. 3. Curves FGW1 and FGW2 are calculated with minimum ($E = 43$ MPa, $h = 3$ cm) and maximum ($E = 250$ MPa, $h = 5$ cm) values of the elastic modulus and ice thickness measured during the tests, and a water depth of $H = 2.5$ m, equal to the water depth in HSVA tank. The gray rectangle shows the region where most of the tests were performed. Figure 3 shows that ice elasticity is not important for wave dispersion with 3cm thick ice when the wave frequency is smaller than 0.7 Hz. The deep-water approximation is valid throughout, since $kH > 2.5$ inside the gray region in Fig. 3.

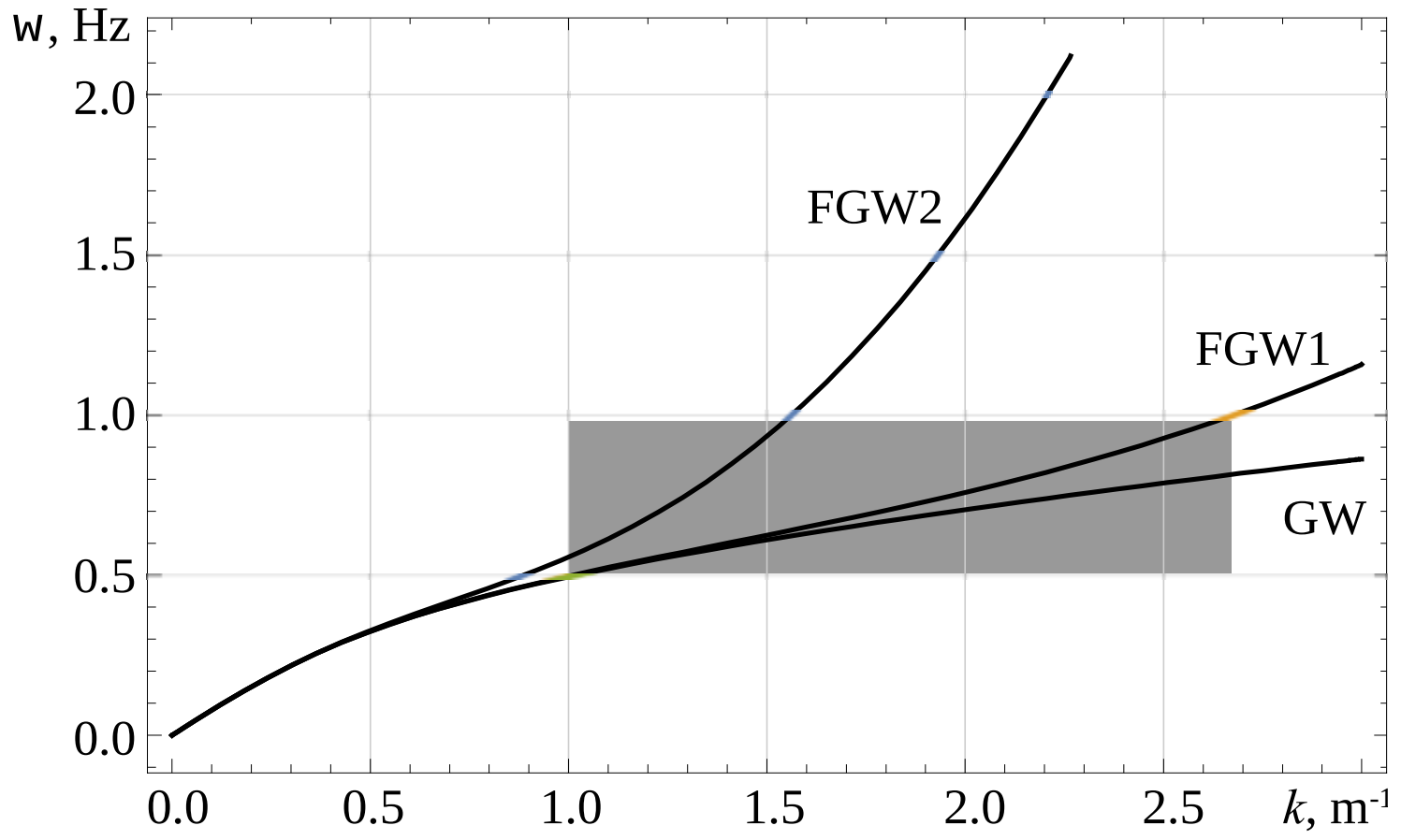

Figure 3. Dispersion equations of flexural gravity waves (FGW1,2) and gravity waves (GW).

Further, the phase velocity of gravity waves in deep water is used for the calculation of $V = g/\omega$, where $\omega$ is the wave frequency in rad/s, in formula (1). The Froude number, calculated with the formula $Fr = \omega^{-1}\sqrt{g/h}$, changes from 2.2 to 5.7 in the tests described here. From dispersion equation (1) it follows that the number $Dk^4$ gives an estimate of the influence of elasticity on wave dispersion (instead of the Cauchy number). We consider the range of wave frequencies where $Dk^4 \approx 1$ (i.e. wave lengths are not very short) so that we can ignore the influence of the gravity force. Expressing the wave number from the dispersion equation for gravity waves in the deep-water approximation ($\omega^2 = gk$) we find that the number $\alpha_{fg} = Eh^3\omega^8/(12\rho_w g^5)$ can be used instead of $Dk^4$. Numerical values of the dimensionless coefficients $E/\sigma_f$ and $\alpha_{fg}$ are shown in Table 4 for the ice characteristics in TGI and TGII.

Table 4. Dimensionless numbers characterizing the bending failure of ice and the influence of elasticity on wave dispersion in TGI and TGII.

| | TGI | TGII | |
|---|---|---|---|
| $E/\sigma_f \cdot 10^{-3}$ | 0.47 | 2.1 | 2.8 |
| $\alpha_{fg}$ | 0.01-2.6 | 0.27-70 | 0.26-68 |

The effective elastic modulus of sea ice measured in full-scale tests with cantilever beams is $E = 1-2$ GPa, and flexural strength is $\sigma_f \approx 0.3$ MPa (Marchenko et al, 2017). Thus, the ratio $E/\sigma_f \cdot 10^{-3} = 3.3-6.6$ is higher in full scale tests than it was TGI and TGII. Figure 4 shows dimensionless numbers $Fr$ and $\alpha_{fg}$ calculated with the characteristics of natural sea ice and natural wind waves and swell. One can see that similarity by Froude number (2.2-5.7 in our tests) can be reached for wind waves with frequency 0.2 Hz (5 s period), swell or local waves with frequency 0.1 Hz (10 s period) in relatively thick ice ($h > 1$m), and almost reached for low frequency swell in thick ice (30 s period and 0.033 Hz frequency). Similarity by $\alpha_{fg}$ (0.01-70 in our tests) can be reached only for waves with frequencies close to or higher than 0.1 Hz (10 s period) propagating in relatively thick ice. The ratio $\rho_i / \rho_w$ is similar for the model and natural ice. Similarity by the number $\alpha_{\sigma h} = \sigma_f / \rho_w g h$, characterizing the influence of hydrostatic pressure on bending failure, is not fulfilled, since $\sigma_f / \sigma_{f,\exp} \approx 3$, while $h / h_{\exp} > 10$.

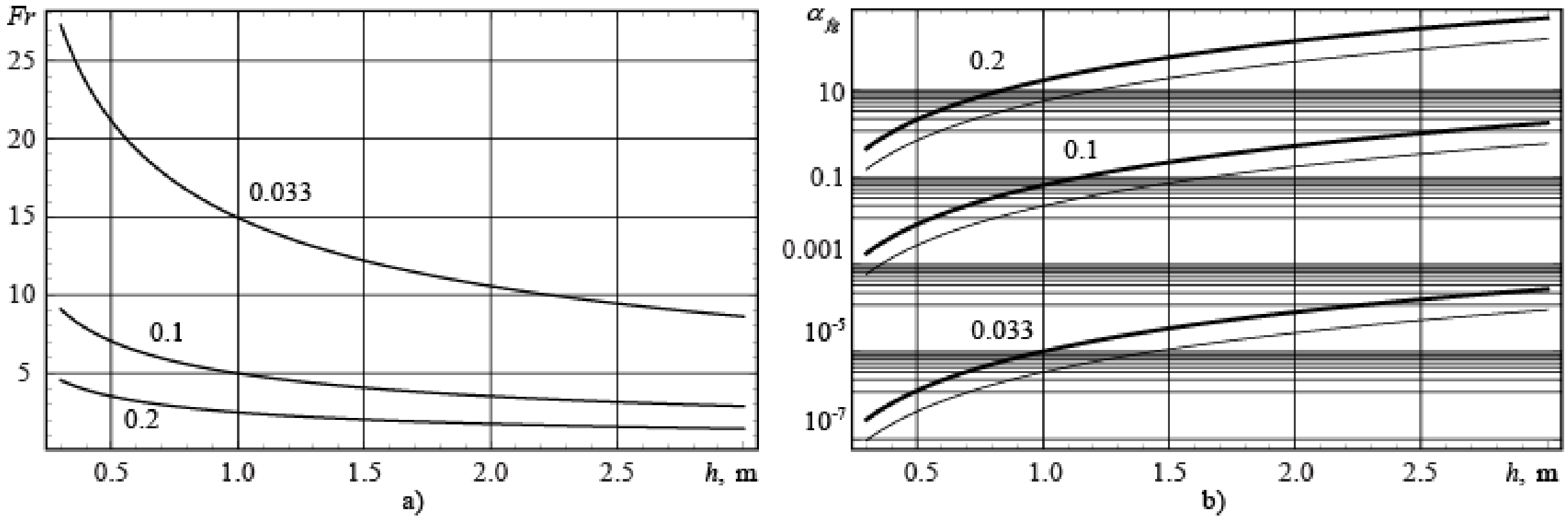

Figure 4. Dimensionless numbers $Fr$ (a) and $\alpha_{fg}$ (b) versus ice thickness, calculated with the characteristics of natural sea ice. Wave frequency (Hz) is marked on the individual curves. Thick and thin lines in (b) are constructed with $E = 3$ GPa and $E = 1$ GPa respectively.

The amplitude of the wave-induced velocity of surface water particles, calculated with standard formulae following from the potential theory of surface waves with small amplitude, equals $\omega a$. The wave amplitudes varied within 0.5 – 1.5 cm, and the wave frequencies varied within 0.5-6 rad/s in the experiment. Therefore, the velocity amplitude is estimated as varying from 1.5 cm/s to 10 cm/s in the tests. At full scale, the velocity amplitude is estimated in the same range when the wave frequency is of about 0.6 rad/s and wave amplitude is of about 10 cm. The decay distance of wave induced motion in the vertical direction is given by $k^{-1}$. According to Fig. 3, this decay distance extends below the ice by 0.4-1 m in the experiment, which excludes any influence of the tank bottom on the waves.

## 4. Observed physical mechanisms of wave damping and ice failure

During the experiment, several factors affecting wave propagation below the ice were observed, including: floods on ice surface (F), perforation of the ice edge (PIE), formation of cracks (CF), pumping of brine through non-through cracks (BP), and production of slush between floes (SP). Figure 5 shows the transformation of the ice edge under wave action during one day of the experiment on 17.01.2018, when 12 tests with duration of 10 minutes were performed. The wave maker was programmed to produced waves of 1 cm height with frequencies varyied from 0.6 Hz to 1.5 Hz.

Increasing the wave frequency caused stronger flooding on the ice edge, which in turn influenced its perforation (Fig. 5b). A non-through crack was recognized along the flood boundary during seventh test (T17_07). By “non-through crack”, here and elsewhere in this paper, we mean a crack which does not split a floe into two. A non-through crack may or may not extend vertically through the ice; it does not extend horizontally through the entire floe. Non-through cracks were monitored by eye and occasionally by hand. Periodic crack opening (with the wave period) was visible at the surface. Eventually, the crack went through the entire floe and an ice band was disconnected from the ice sheet in the twelfth and final test (Fig. 5c).

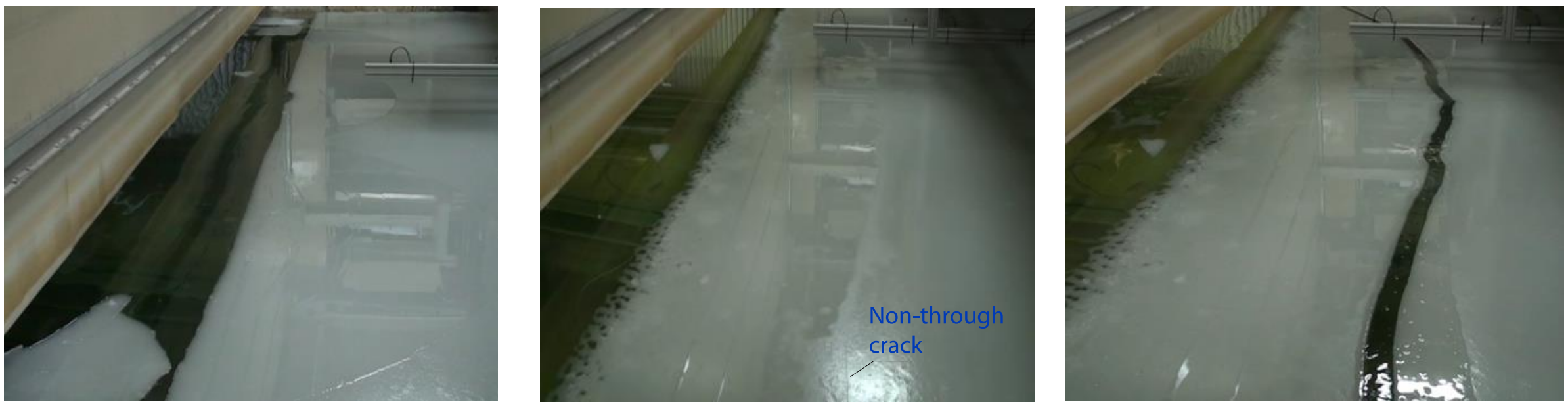

a) b) c)

Figure 5. (a) Flood on the ice surface due to wave action on the ice edge (T17_10); (b) perforated ice edge (T17_07); (c) formation of crack separating an ice block from the ice edge (T17_07_02).

Six tests with larger wave heights, varying from 15 mm to 60 mm, were performed on the next day (18.01.2018). Three tests were performed with wave frequency of 0.7 Hz and three tests with wave frequency of 1 Hz. The ice cover was partially destroyed, breaking into floes with sizes smaller than 1m around main carrier by $x$<30 m (Fig. 6). Video footage shows floe interactions by collisions during wave propagation. Repeated collisions caused the formation of floods, slush between floes, and smoothing of sharp corners of the floes. Floe collisions also caused displacements and rotations of the floes. These motions produced turbulence between floes which was observed visually.

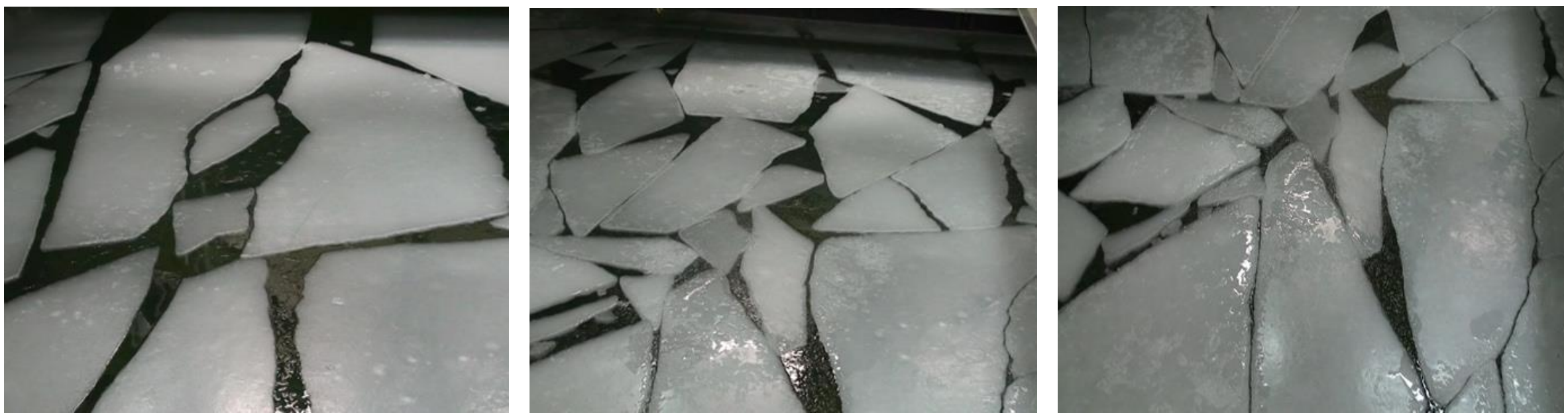

a) b) c)

Figure 6. (a) Beginning of slush formation after ice break up in the test T18_07; (b) the same broken ice in the test T18_10; (c) the same broken ice in the test T18_07_02.

The ice located in the tank with $x > 30$ m remained undestroyed (Fig. 7). A system of several non-through cracks similar to shown in Fig. 5b, extending across the tank and providing effective wave damping, was discovered. These cracks were clearly visible from a distance, due to their cyclic opening and closing during wave propagation (see, e.g., cracks 1 and 2 in Fig. 8). Cycling pumping of the brine was also clearly visible from when the cracks were observed up close (Fig. 9). It was not possible to locate cracks at the bottom of the ice by manual inspection near the tank wall, while they were clearly visible at the upper ice surface, where the cracks pass through surface granular layer of model ice (see Fig. 2). Columnar ice has better permeability, and allows liquid brine to migrate through the ice under the action of the pressure gradient caused by bending deformations of the ice. An outline scheme of the brine pumping caused by periodical compressions and tensions in the surface and bottom layers of the ice (which themselves are caused by bending deformations) is shown in Fig. 10.

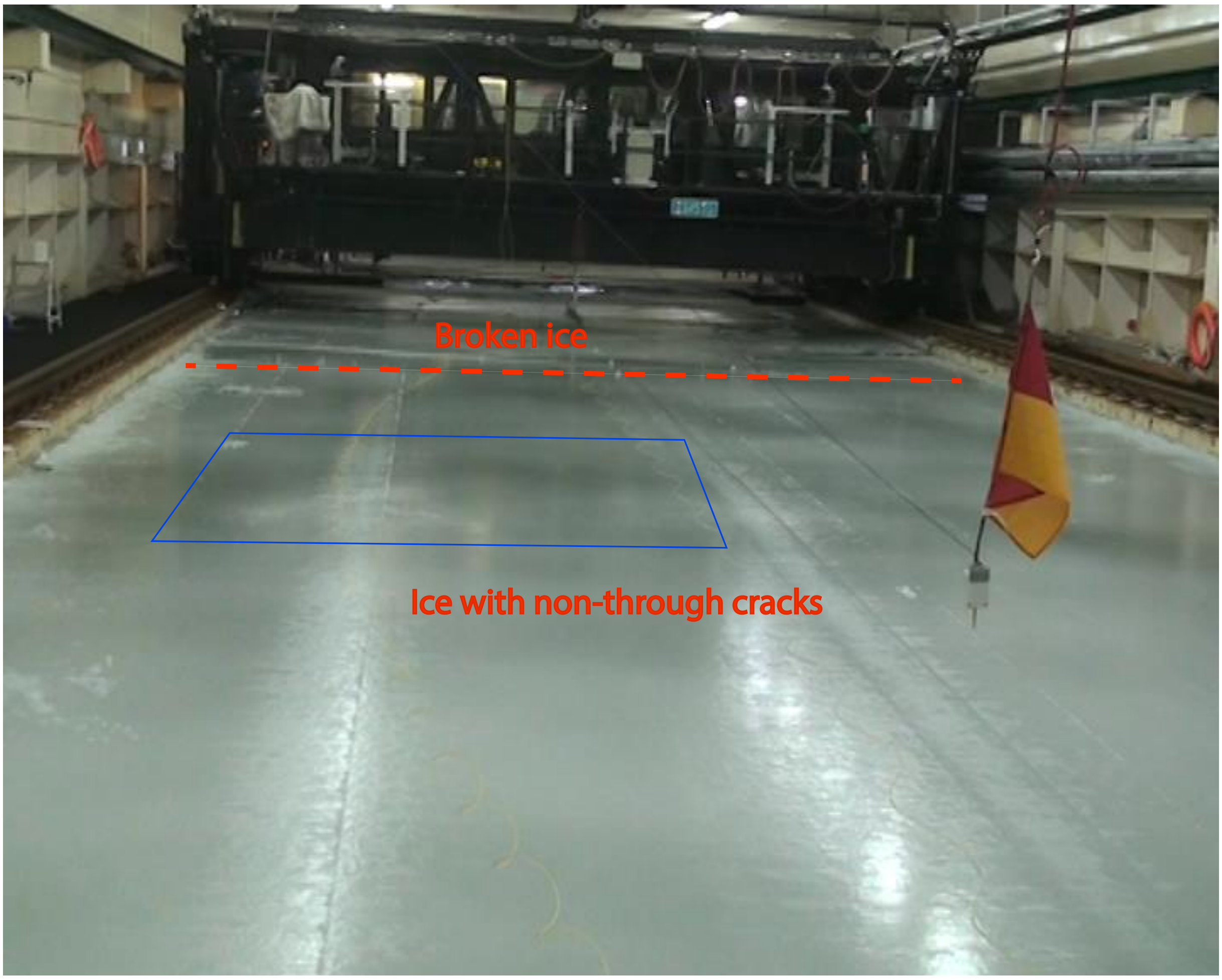

Figure 7. General view of ice in the test T18_07_01.

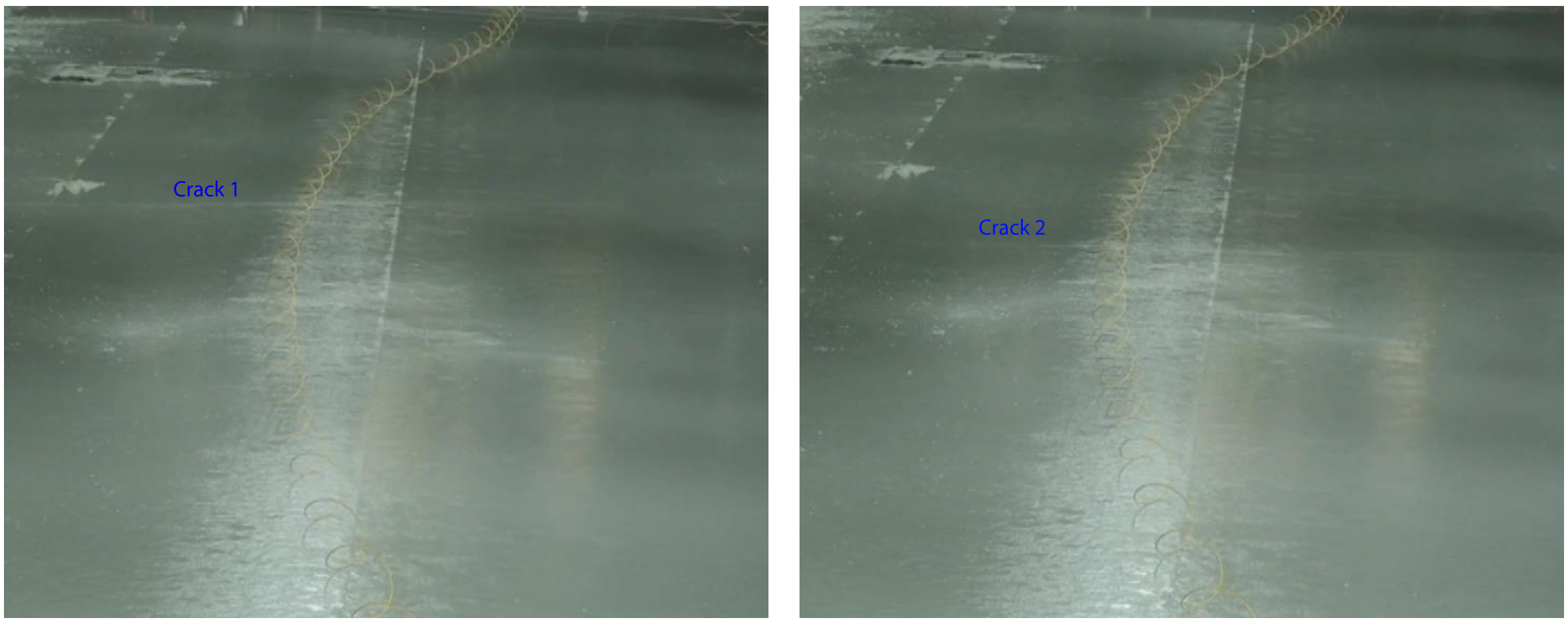

Figure 8. Sequential opening of two non-through cracks located inside blue contour in Fig. 7.

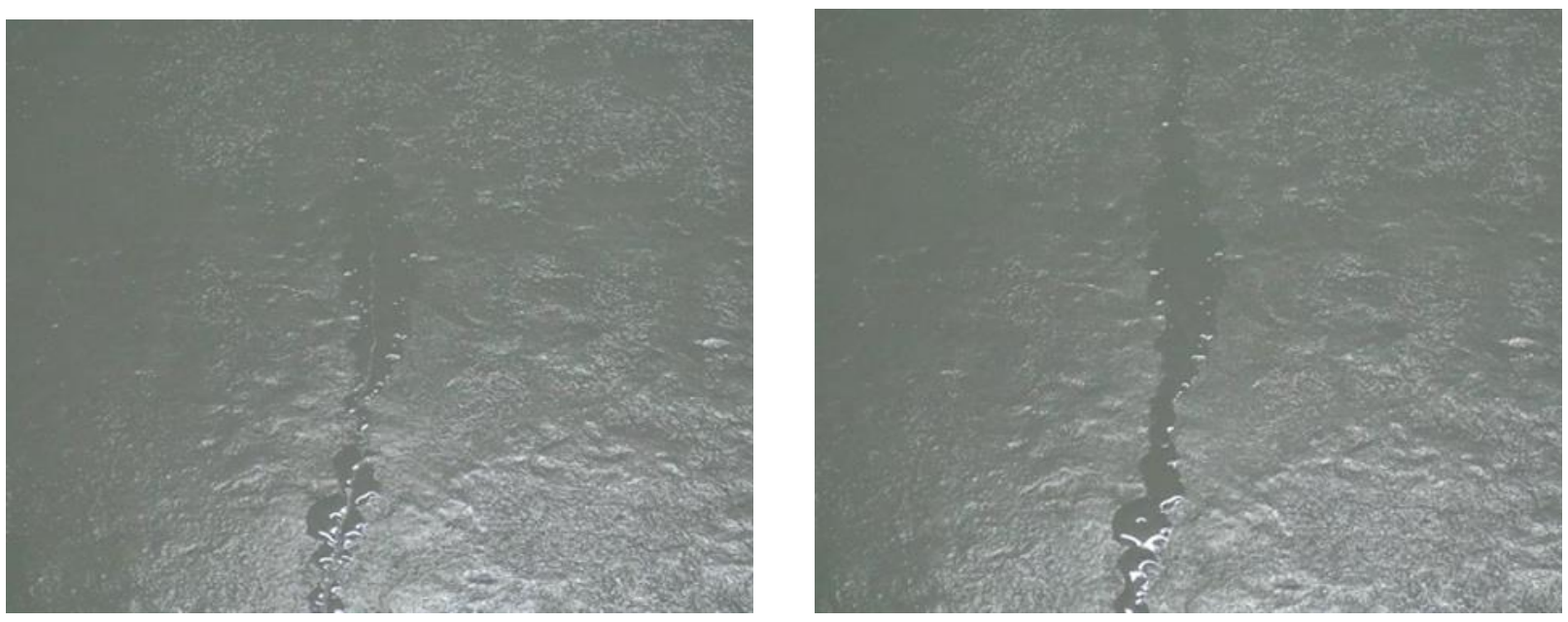

a) b)

Figure 9. Photographs of (a) closed, and (b) open and brine-filled non-through cracks.

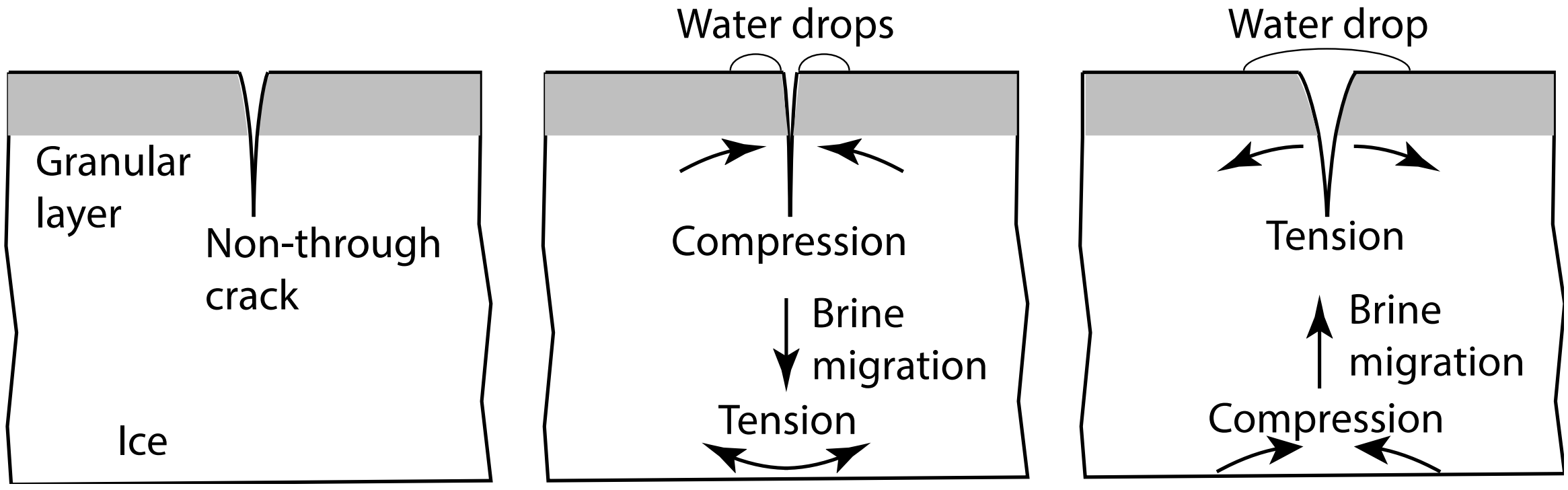

Figure 10. Scheme of brine migration through the non-through crack due to the ice bending.

5. Deployment and selected results of the Qualisys motion capture system

Five markers from a Qualisys–Motion Capture System were used to record ice movements at 5 points (Fig. 11). The data includes the records of the three coordinates of each marker as a function of time, with sampling frequency of 200 Hz. An example of the record of the ice motion along the tank in the test T16_08_mov is shown in Fig. 12a. Figure 12b shows that ice displacements across the tank in the same test had magnitudes of a few millimeters. All marker records during the cyclic-motion experiments show similar motion, corresponding to almost solid body cyclic motion of ice along the tank axis. The amplitude of the motion along the tank axis was not controlled and can vary from cycle to cycle within several tens of centimeters. Figure 13a shows spectrum of the ice displacement in the x-direction in the test T16_08_mov. The spectral maximum marked MF corresponds to the period of 40 s - 50 s of the forced ice drift along the tank.

Figure 13b shows spectrums of the ice displacements in the *y*- and *z*-directions in the test T16_08_mov. Spectral maxima of the *y*- and *z*-displacements at the MF frequency are clearly visible in the figure. The *z*-displacement has, in addition, spectral maxima at natural frequencies of the tank (M1, M2 and M3). The period of the first natural frequency is estimated with Merian's formula as $T_1 = 2L/\sqrt{gH}$, where $L = 70$ m is the tank length and $H = 2.5$ m is the tank depth. Periods of the first three natural frequencies are $T_1 = 28.3$ s, $T_2 = 14.13$ s and $T_3 = 9.43$ s, and their frequencies respectively are $MF1$=0.0353 Hz, $MF2 = 0.0707$ Hz, and $MF3 = 0.106$ Hz. It is of interest that spectral maxima corresponding to the second and the third natural modes are much greater than the spectral maximum of the first natural mode of the vertical oscillation in the *z*-direction. This can be explained by the confinement of the vertical displacement of the ice at the end of the tank.

Figure 14 shows spectrums of the vertical displacement of one marker versus calculated using the data from 3 tests (T16_06, T16_08, and T16_10) with fixed ice (blue lines) and 3 tests (T16_06_mov, T16_08_mov, and T16_10_mov) with moving ice (yellow lines). The numbers 1, 2 and 3 correspond to the order of the test performances. The test order is specified since turbulence may accumulate in the tests with moving ice, which were performed in a raw one after the other. Spectral maxima coincide with the wave frequencies in all the tests. One can see that the cyclic motion of ice causes spreading of the spectrums around the wave frequency and significant reduction of the peak values: that means a reduction of the wave amplitudes. The effect is always stronger in the last tests marked by number 3. This indicates that the spectral spreading and reduction of wave amplitude in the experiments with moving ice are stronger for the waves of higher frequency.

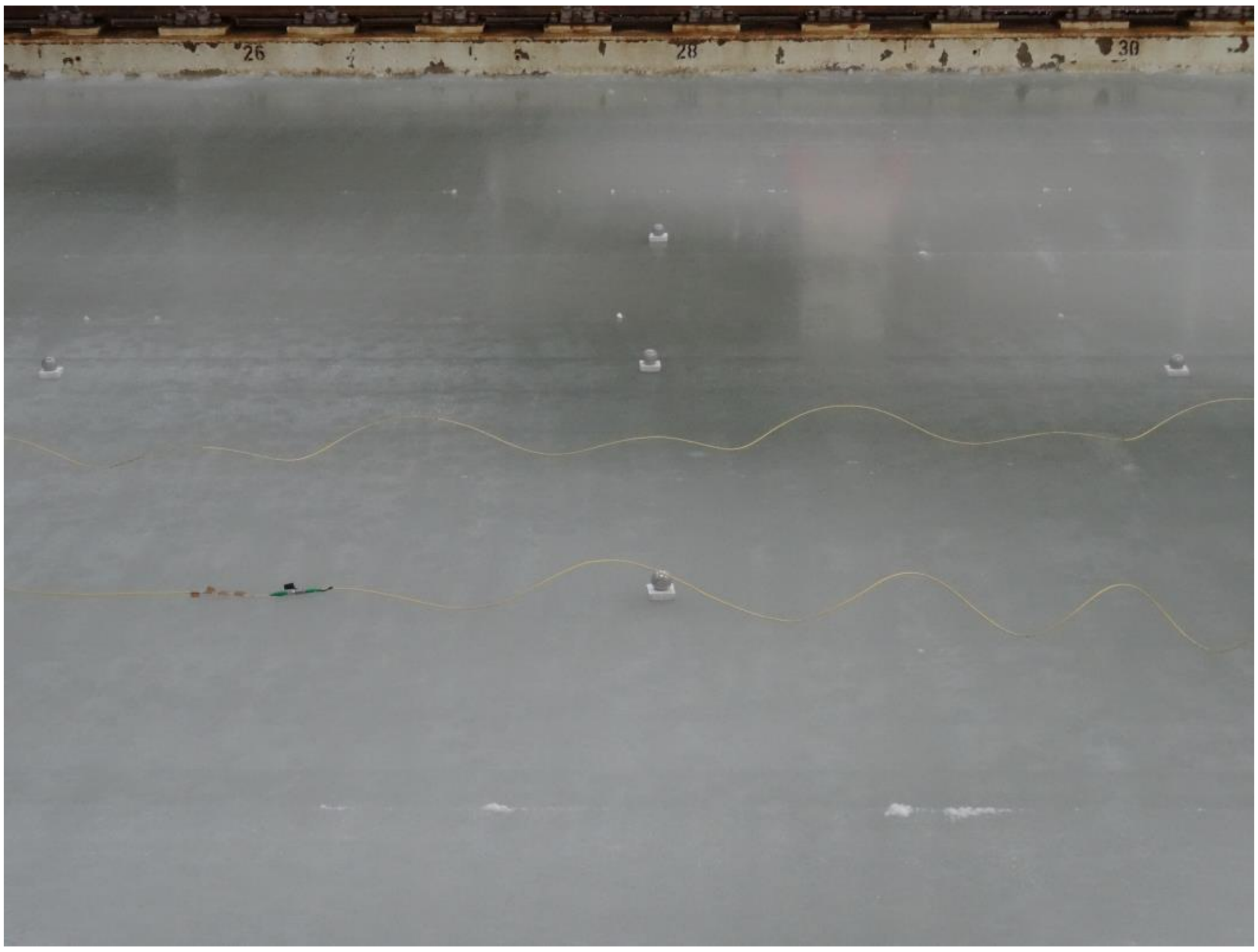

Figure 11. Markers for the Qualisys system on the ice.

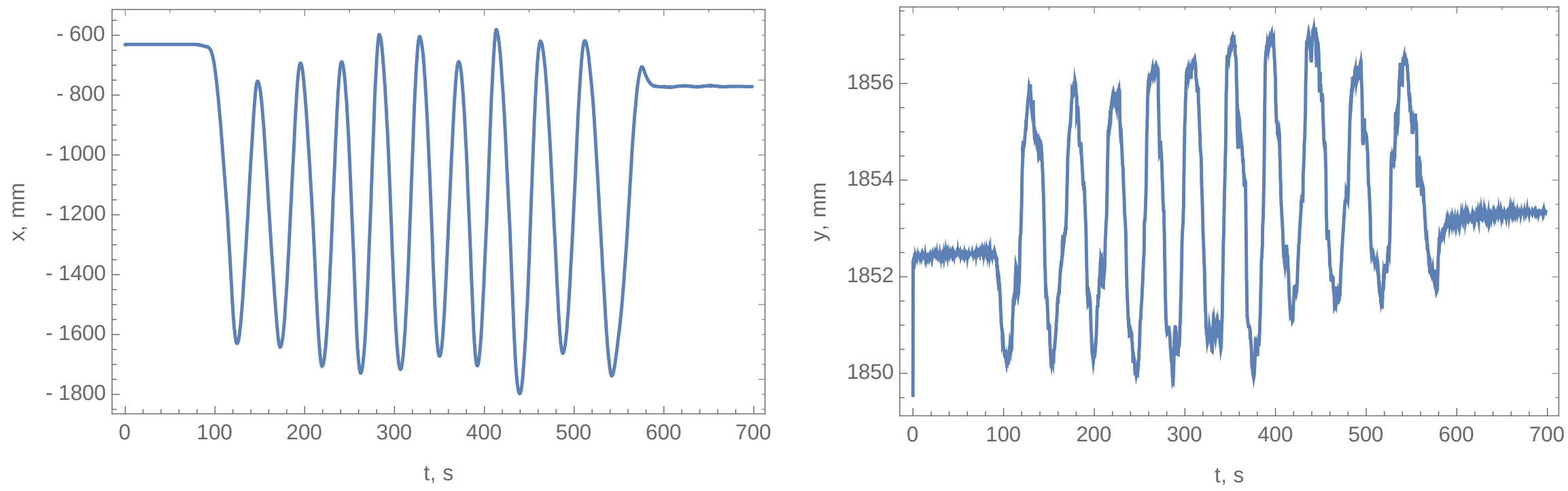

a) b)

Figure 12. Records of the ice marker motion along the tank (a) and across the tank (b) in the test T16_08_mov.

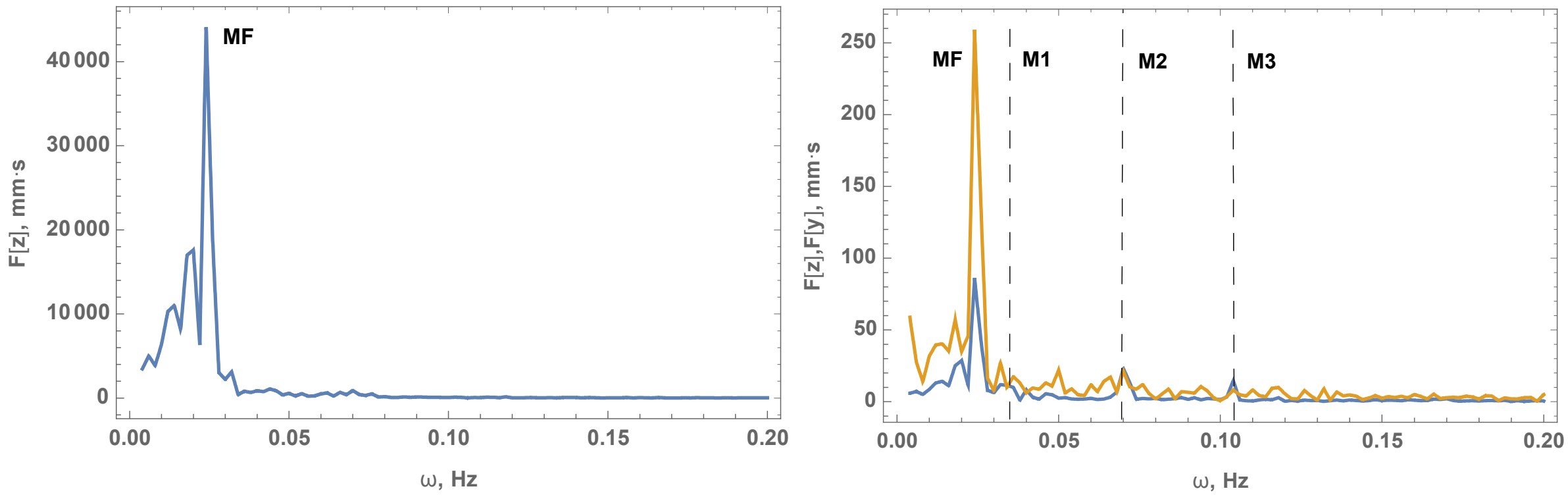

a) b)

Figure 13. Fourier spectrums of ice displacements in the x-direction (a), y-direction (yellow line) and z-direction (blue line) (b) in the test T16_08_mov .

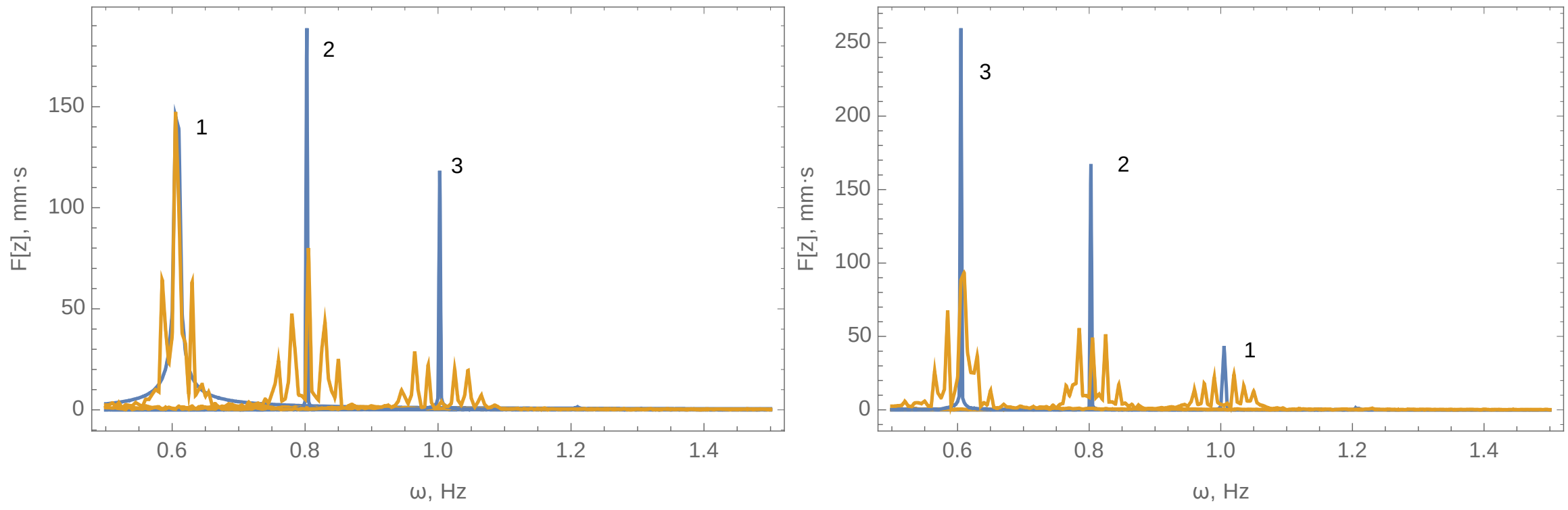

a) b)

Figure 14. Fourier spectrums of vertical oscillations of the ice in the tests with fixed (blue lines) and moving (yellow) ice: (a) T16_06, T16_08, T16_10, T16_06_mov, T16_08_mov, T16_10_mov, (b) T17_06, T17_08, T17_10, T17_06_mov, T17_08_mov, T17_10_mov.

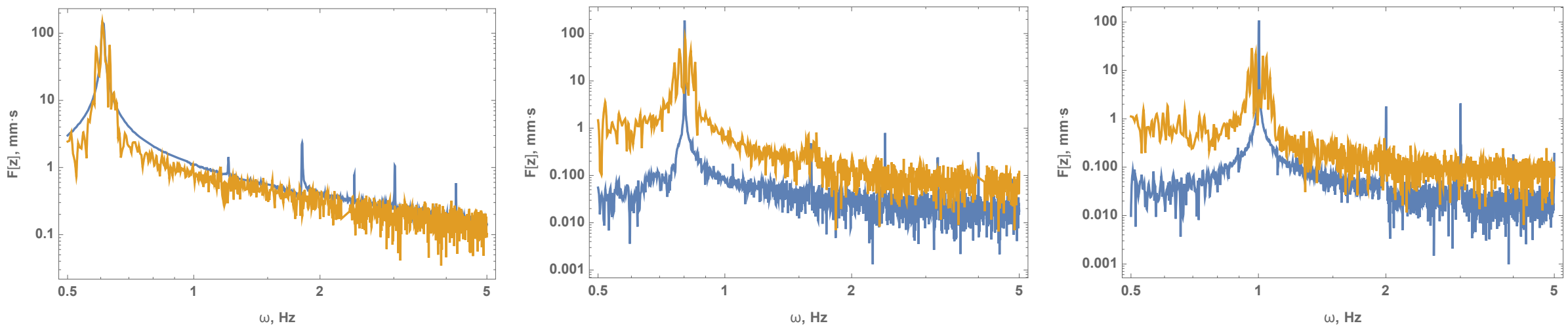

a) b) c)

Figure 15. Fourier spectrums of vertical oscillations of the ice in the tests with fixed (blue lines) and moving (yellow) ice: (a) T16_06 and T16_06_mov, (b) T16_08 and T16_08_mov, (c) T16_10 and T16_10_mov.

Figure 15 shows spectrums of the vertical displacement of one marker versus frequency in wider spectral range. Spectral peaks associated with multiple harmonics of the wave frequencies are clearly visible on the blue lines (corresponding to the tests with fixed ice). They are absent on the yellow lines, corresponding to the tests with moving ice. This is explained by the stronger damping of waves with higher frequency in the experiments with moving ice, which is also clearly visible from Fig.14.

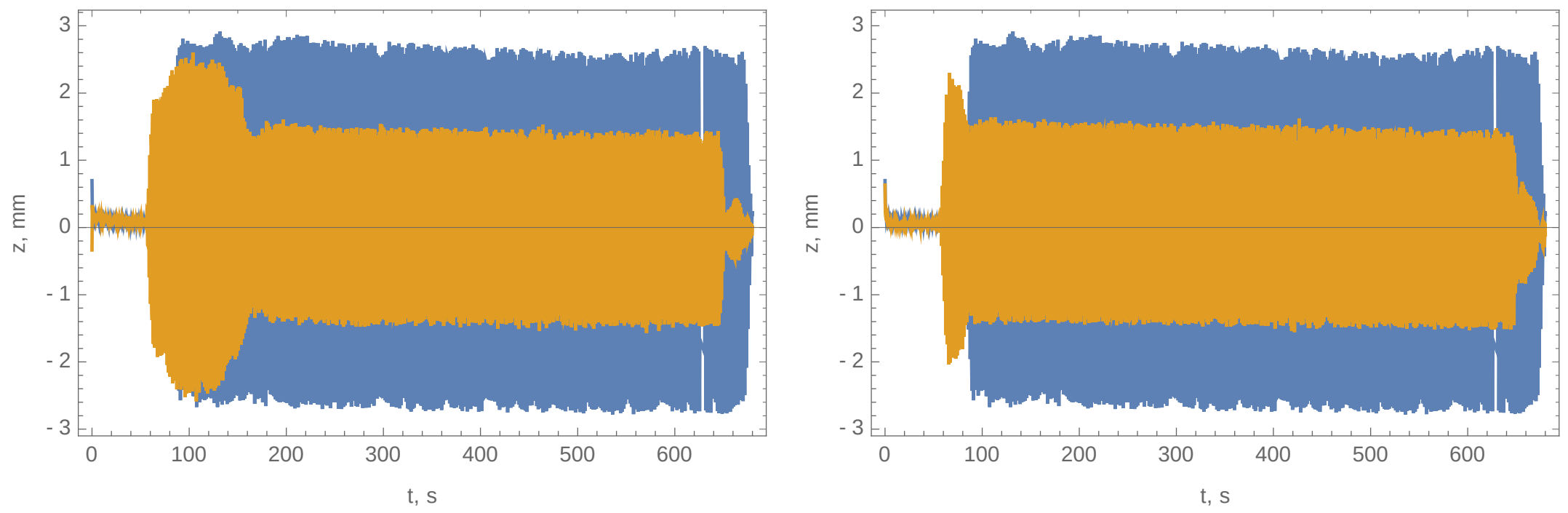

a) b)

Figure 16. Records of the vertical displacement of the ice marker before (blue line) and after (yellow lines) the formation of the non-through crack in the ice. The blue line (a,b) is recorded in the test T17_06, and the yellow lines are recorded in the tests T17_07 (a) and T17_07_01 (b).

Records of the vertical displacement of the ice marker before (blue line) and after (yellow lines) the formation of the non-through crack in the ice are shown in Fig. 16 versus the time. The blue lines in Fig. 16a and Fig. 16b obtained from the test T17_06 are similar. The yellow lnes correspond to the tests T17_07 (a) and T17_07_01 (b). The test T17_06 was performed before the formation of the non-through crack near the ice edge (Fig. 5a), and two tests T17_07 and T17_07_01 were performed after the formation of the non-through crack (Fig.5b). Note that Fig. 16a and Fig.16b show the same effect of very significant reduction of the wave height due to the action of the non-through crack on incoming waves occurring across two sets of experiments, which gives evidence that this is a repeatable phenomenon. During the last two tests performed in the same day this crack passed through the ice (Fig. 5c).

## 6. Deployment and selected results of ADCP and ADV

Point measurements of the water velocity were taken with a 3D current meter (Nortek Vector), and measurements of the vertical profile of 3D water velocity were taken with three current profilers (Nortek Signature 1000). The scheme of the ADV and ADCP deployments is shown in Fig. 17. These results are grouped together, and similar short names - ADV for the 3D current meter and ADCP for the current profiler - are used since the measurement principle for both of these sensors is based on the Doppler effect. The diameter of the ADV sampling volume was 8.6 mm and sampling rate was 64 Hz. The ADV was mounted in an upwards-looking position on a frame standing on the bottom of the tank. The distance between the ADV probe and wave maker was 39 m, and the distance between the ADV probe and the tank wall was 2.5 m. The sampling volume was ~15 cm distant from the ice bottom, and the distance between the ADV transducer and the sampling volume was also 15 cm. The ADCP probes were placed in the middle of the tank, at the bottom, in upwards-looking positions, in three locations. The locations 1-3 are 10 m, 17 m and 39 m distant from the wavemaker. Location 1 is in the open water region. The vertical size of the ADCP probes is 30 cm. The sampling rate of ADCP measurements was 8 Hz. Each sample includes values of 3D velocity in 17 cells averaged over the cell size of 20 cm. Only first 10 cells are located in the water layer below the ice. The first cell is separated from the ADCP head by the blanking distance of 10 cm. Recording of the ADV and ADCP data was performed in online mode by cables connected to a laptop.

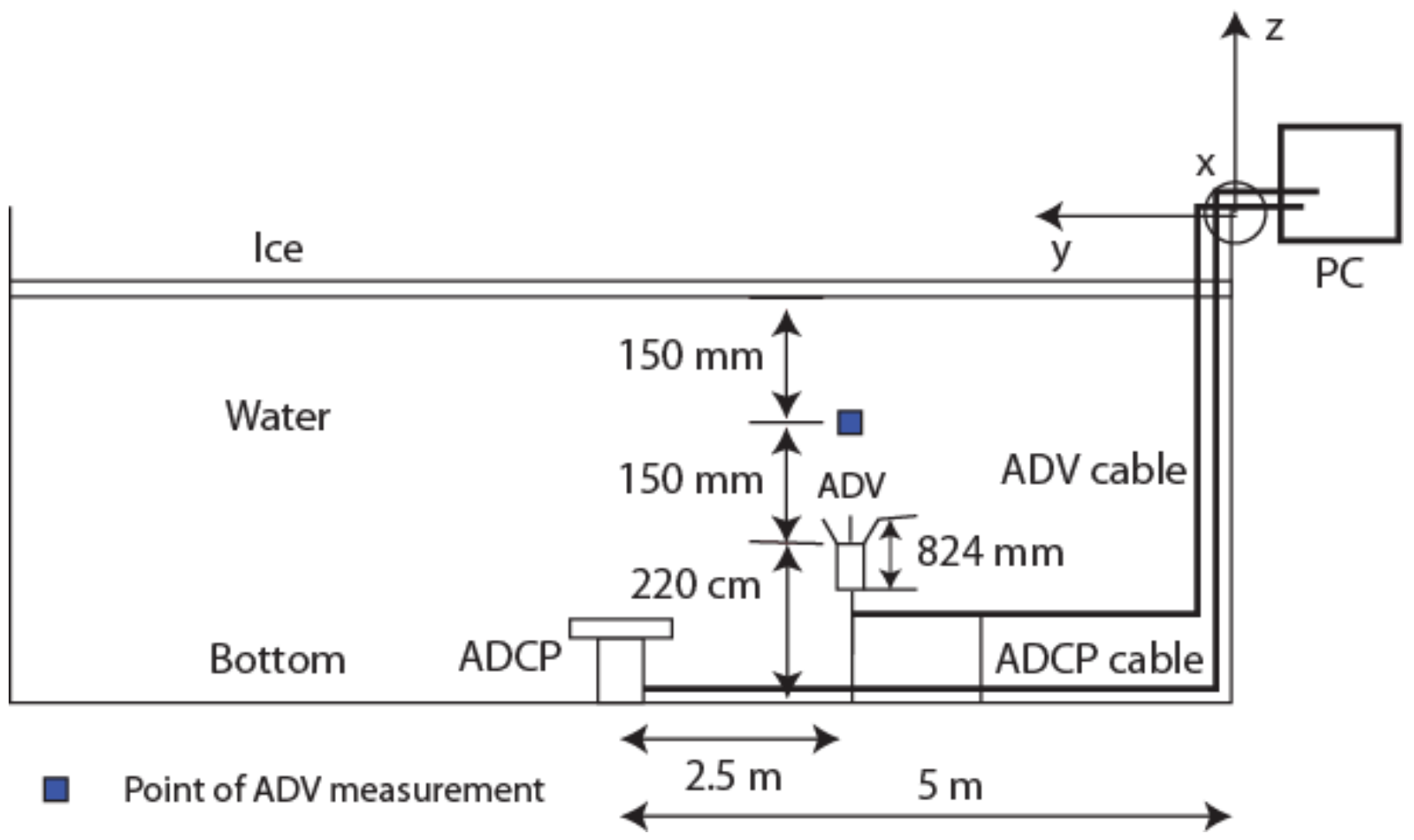

Figure 17. Scheme of the ADV and ADCP deployments.

Figures 18 and 19 show results of data analysis of the ADV and ADCP records in the tests with fixed ice (T16_08) and moving ice (T16_08_mov). The elastic modulus of ice varied from 88 MPa (before the tests) to 126 MPa (after the tests) in the day when these tests were performed. This corresponds to wavelengths of 3.93 m and 4.15 m. Figures 18a and 18b show the spectrums of water velocity in the $x$- and $z$- directions. There are spectral peaks at the frequency of about 0.8 Hz in Fig. 18a. The mean correlation of the ADV records is around 80%, and mean SNR is 12 db. The yellow lines show slightly higher energy stored in low frequency oscillations in the experiments with fixed ice. Figure 19a shows ADCP records of the vertical velocity averaged over a 20 cm water layer located at 30 cm distance below the ice. Amplitudes of the velocities are higher in the experiment when the ice was moving along the tank. Comparison of the spectrums shown in Fig. 19b shows a higher energy of velocity fluctuations in the experiment with moving ice. There are spectral peaks at the wavemaker frequency of about 0.8 Hz, and on the multiple frequencies in Fig. 19a. A comparison of the blue lines in Fig. 18a and Fig. 19b shows energy flux from high to lower frequencies with increasing distance from the ice bottom in the experiments with moving ice.

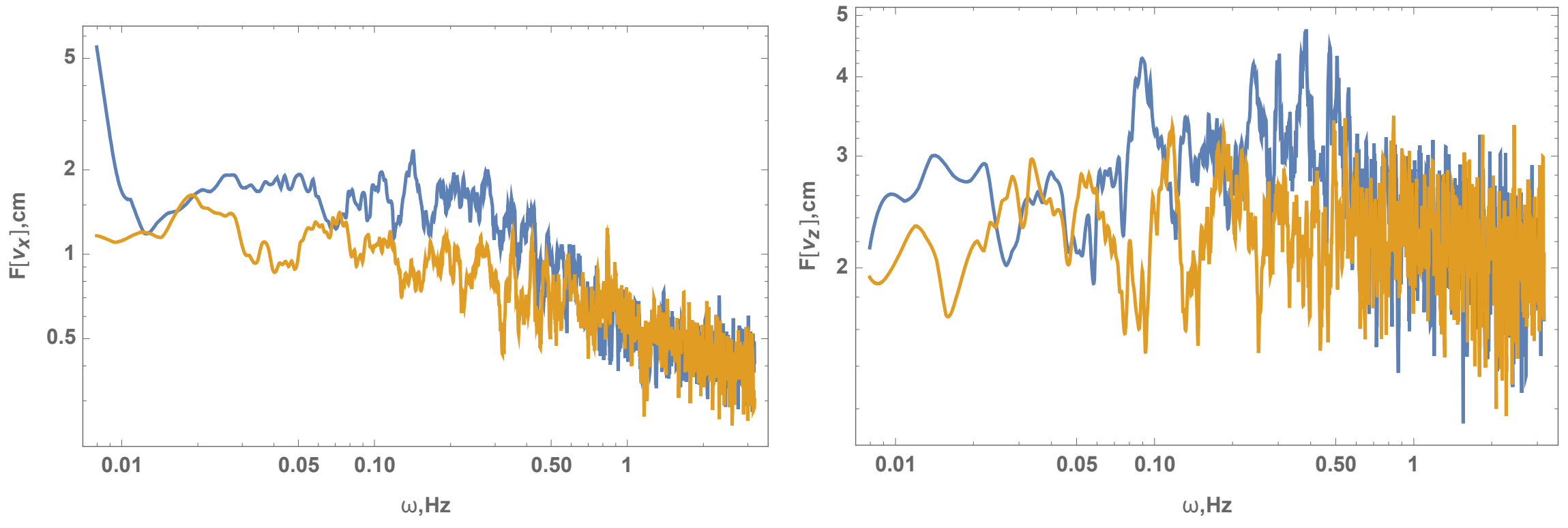

a) b)

Figure 18. Spectrums of the horizontal (a) and vertical (b) water velocities measured by the ADV. Blue and yellow lines correspond to the experiments with fixed and cyclically moving ice.

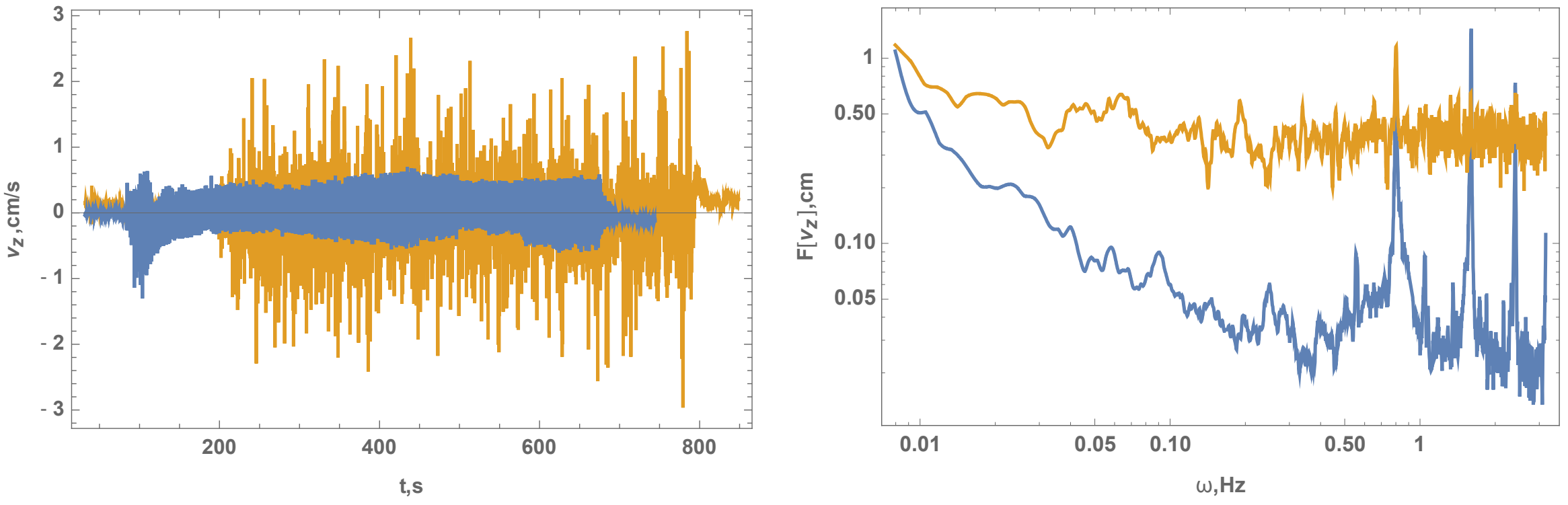

a) b)

Figure 19. ADCP records of the vertical water velocity versus the time in cell 9 (a). Spectrums of the vertical velocities (b). Blue and yellow lines correspond to the tests with moving ice and fixed ice.

The quality of the ADV and ADCP data is limited by a low amount of seeding material in the tank. Sidelobe interference may influence the recorded signals of inclined beams on the ADCP in two cells below the ice, i.e. records of the four beams can't be used for the calculation of the velocity components in two cells (extended from 10 cm to 30 cm and from 30 cm to 50 cm from the ice bottom in downward direction).

7. Deployment and selected results of FBG sensors

An FBG thermistor string (with 12 distributed thermistors distributed inside a metal tube) and an FBG strain sensor are shown in Fig. 20. Neighbouring thermistors are spaced 1cm apart (Fig. 21). Typical strain resolution for FBG systems is 1 μstrain ($10^{-6}$) or better, and the accuracy is typically 5 μstrain. The FBG temperature measurement system's nominal resolution and accuracy in the experiment was 0.08° C and 0.4° C, respectively. The variation ($\Delta\lambda$) of the peak wavelength caused by the extension ($\Delta L / L$) and the change of the temperature ($\Delta T$) of the sensor is described by the equation

$$\frac{\Delta\lambda}{\lambda} = GF \cdot \frac{\Delta L}{L} + TK \cdot \Delta T \ , \tag{3}$$

where the gauge factor $GF = 0.719$ and a linear temperature coefficient $TK = 5.5 \cdot 10^{-6}$ are the constants obtained from a calibration cycle for the FBG sensors in standard SMF fiber, within a temperature range from -20° C to 0° C. The variation of the peak wavelength $\Delta\lambda$ is measured with a spectrometer that receives the reflected signal from the FBG sensor. To calculate strain ($\Delta L / L$) using formula (3) it is necessary to measure the temperature change ($\Delta T$) at the strain sensor's position, in order to compensate for thermal expansion effects. The temperature measurements can easily be performed with another FBG sensor protected from mechanical deformation, or alternatively with a thermometer.

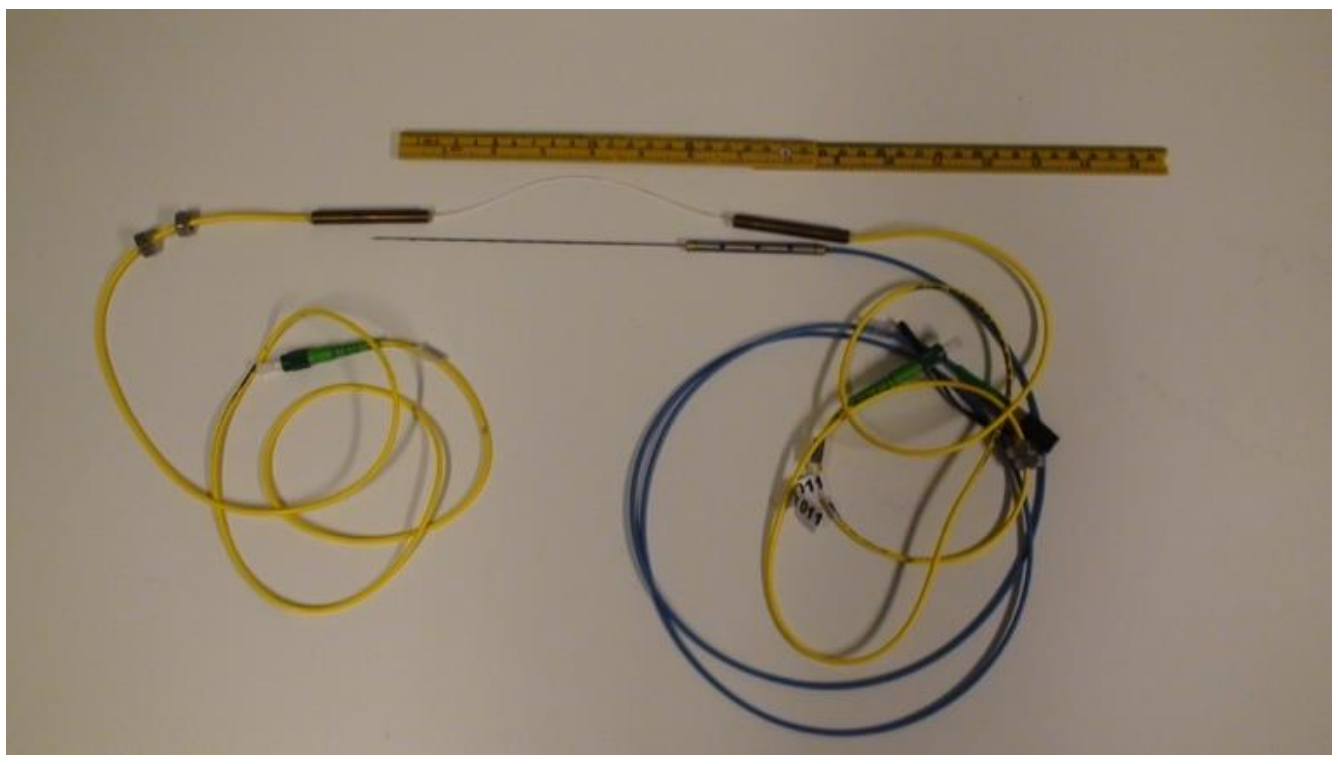

Figure 20. An FBG thermistor string and strain sensor.

The FBG sensors were used in the experiments to measure in-plane strains in the ice (excited during propagation of surface gravity waves below the ice) and to record a vertical profile of the temperature (in the water layer below the ice, in the ice and above the ice) over a distance of 12 cm (with spatial resolution of 1 cm). A schematic of the installation of the strain and temperature sensors is shown in Fig. 21. Each strain sensor (FBGS sensor) measures strain (FBG strain) between two points where the fiber is fixed to bolts, which in turn connect the working length of the fiber (including the FBGS sensor) to the fiber which transmits optical signal. The bolts are fixed onto brackets with nuts and washers, and each bracket is mounted on the ice with four screws. It is evident that FBG strain consists of a sum of the in-plane strain in the ice and the strain due to the bracket tilts caused by ice bending. Four FBGS sensors were deployed to measure longitudinal ($x$-direction) strains in the ice at distances (x-direction) of 19 m (2 sensors) and 50 m (2 sensors) from the ice edge. Another four FBGS sensors measured strains in the transversal direction to the tank axis ($y$-direction) in similar locations. Two FBG temperature strings (FBGT sensors) were supported by foam plastic holders so that 3 thermistors were above the ice surface. These FBGT sensors were then placed inside holes of 2 mm diameter drilled through the ice. The diameter of FBGT sensors is slightly smaller than 2 mm. Therefore, FBGT sensors were tightly held inside the holes without visible gaps. Photographs of the sensors, installed in position, are shown in Fig. 22.

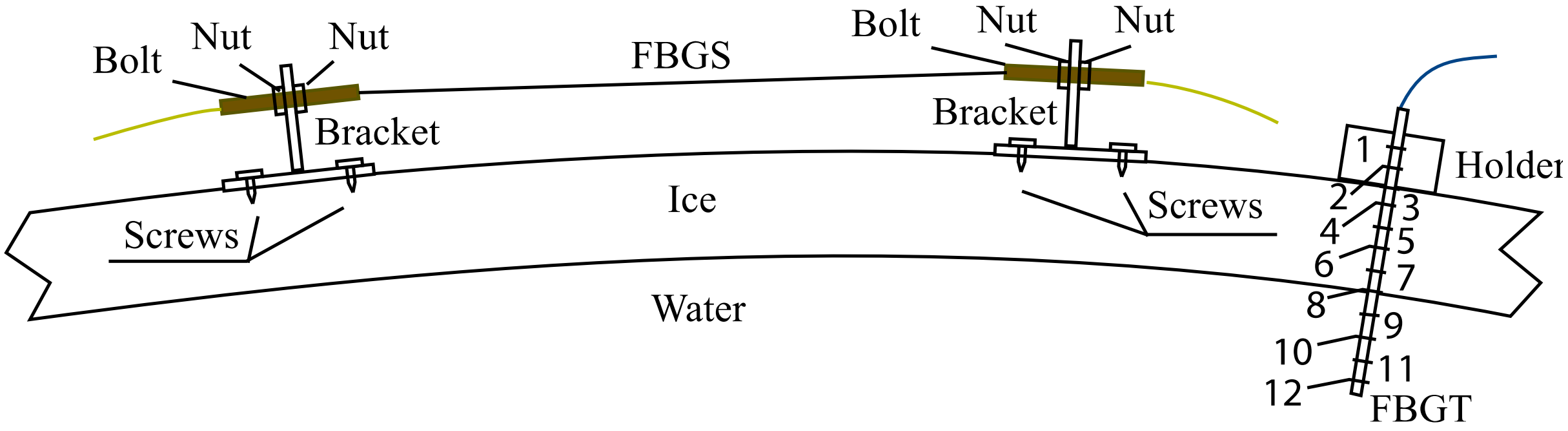

Figure 21. Schematic of the installation of an FBG strain sensor (FBGS) and temperature string (FBGT) on the ice.

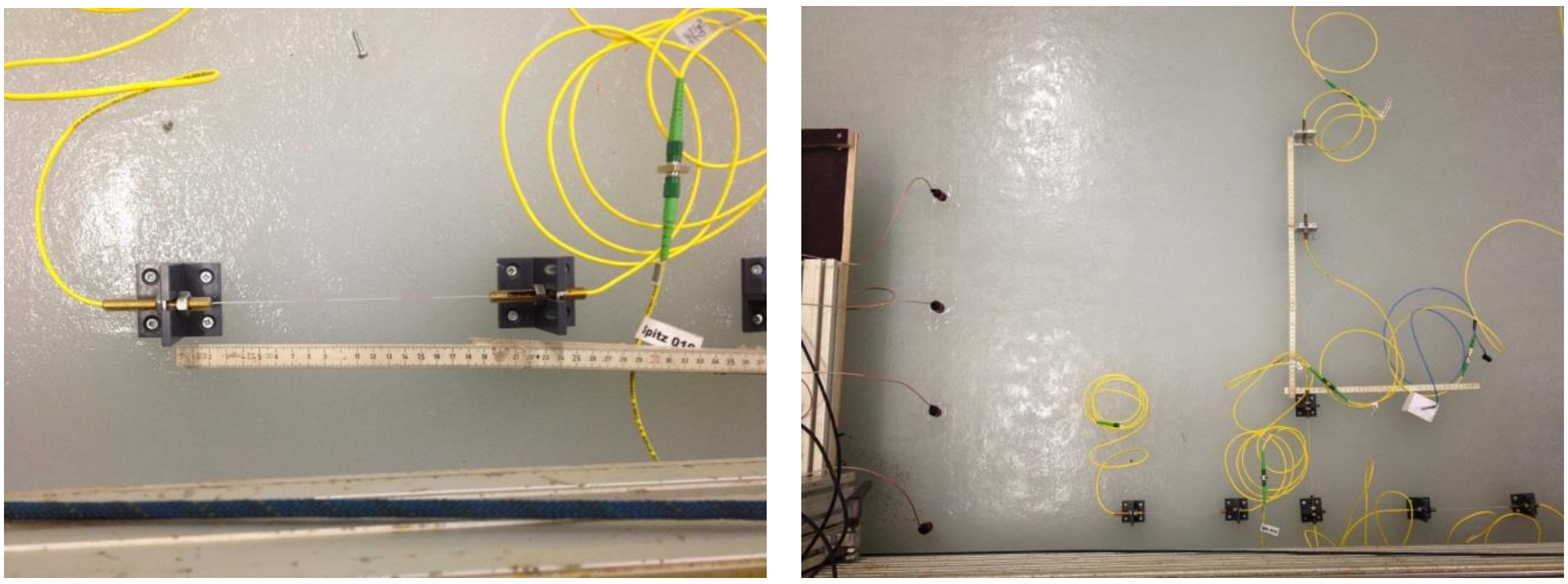

Figure 22. Mounting of FBG strain sensor on the ice (a). Four FBG strain sensors (FBGS) and thermistor string (FBGT) mounted on the ice near AE sensors (b).

Measurements of strain and temperature were recorded at a frequency of 40 Hz. Strain records show a periodic dependence on time, with a dominant period equal to the wave period. In all tests the amplitudes of longitudinal strains were much higher than the amplitudes of strains in the transverse direction. Figure 23 shows FBGS records made during the tests with fixed ice (T16_08) and moving ice (T16_08_mov). The low frequency oscillations/modulations of strains in Fig. 23 c,d corresponds to the period of cyclic motions of the ice (Fig. 12). The strain amplitudes at 50 m distance from the ice edge (FNGS5 in Fig. 1b) are lower than the strain amplitudes at 19 m distance from the ice edge (FBGS1 in Fig. 1b) because of the wave damping. The wave damping is stronger in the tests with moving ice. The linear increase in measured strains (shown in Fig. 23 a,c by black lines) corresponds to a stretching of the ice sheet under the wave action. The resulting stretching of the ice is also greater in the experiments with moving ice in comparison with the fixed ice experiment.

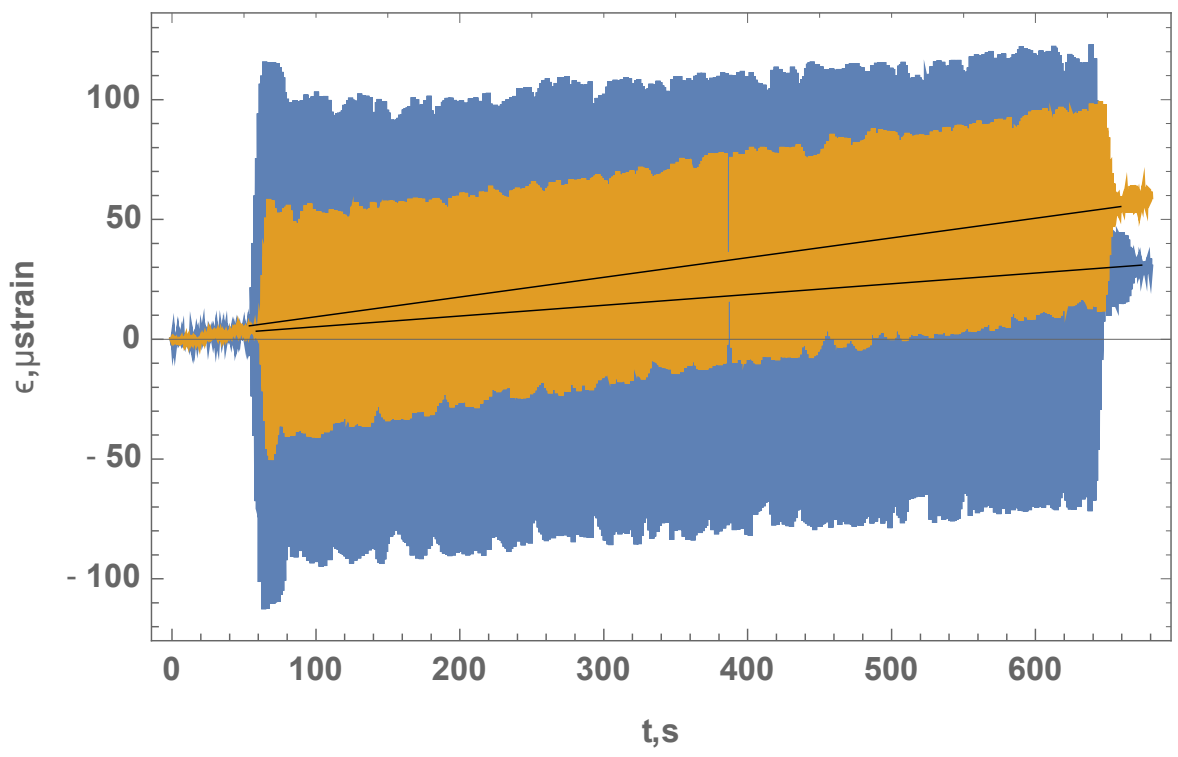

a)

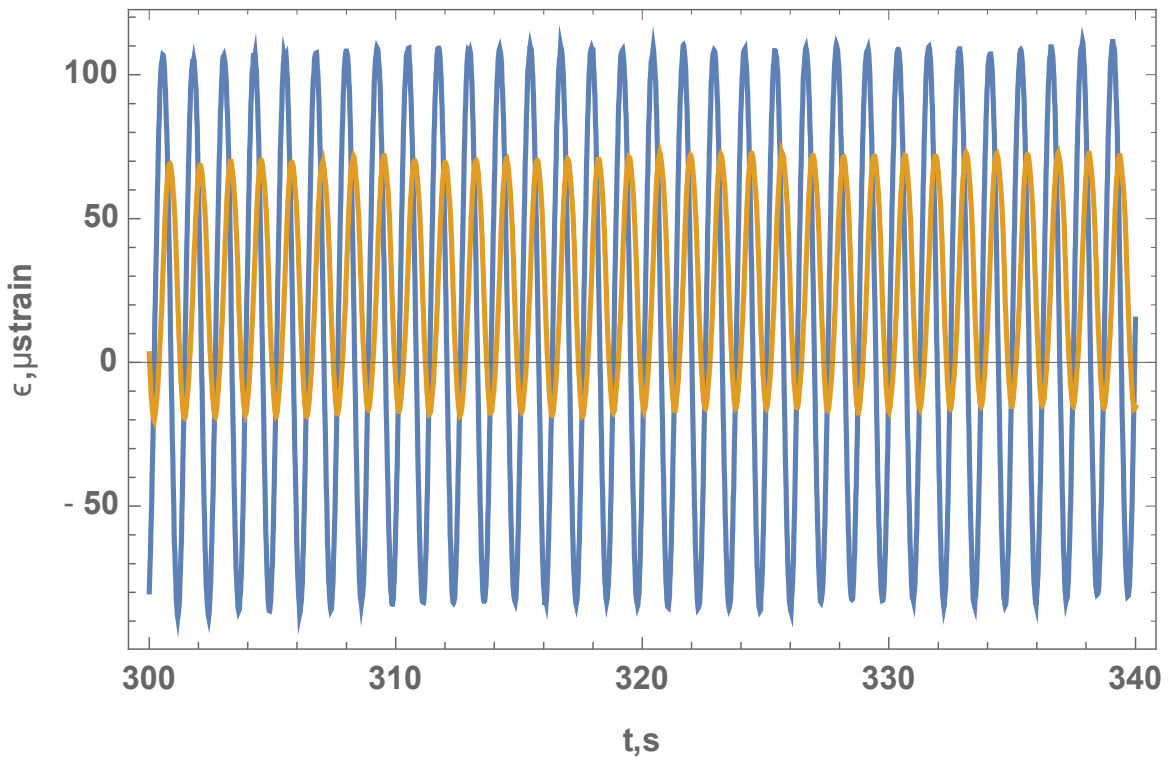

b)

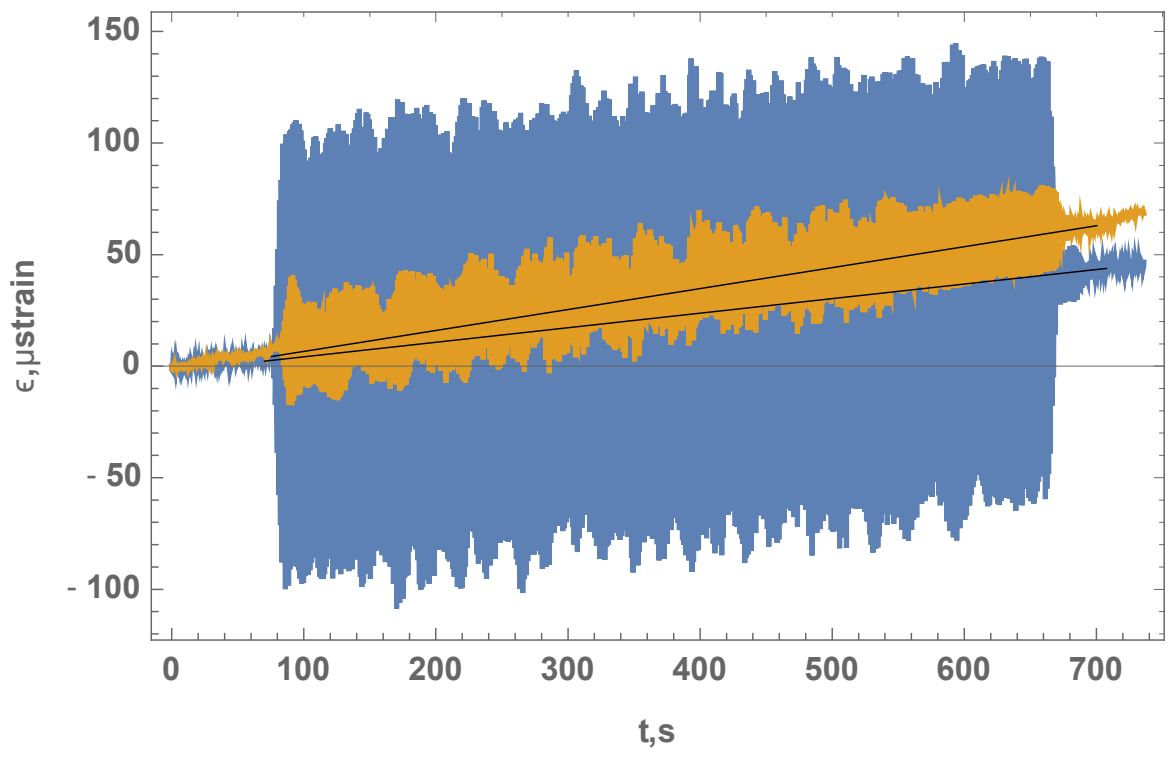

c)

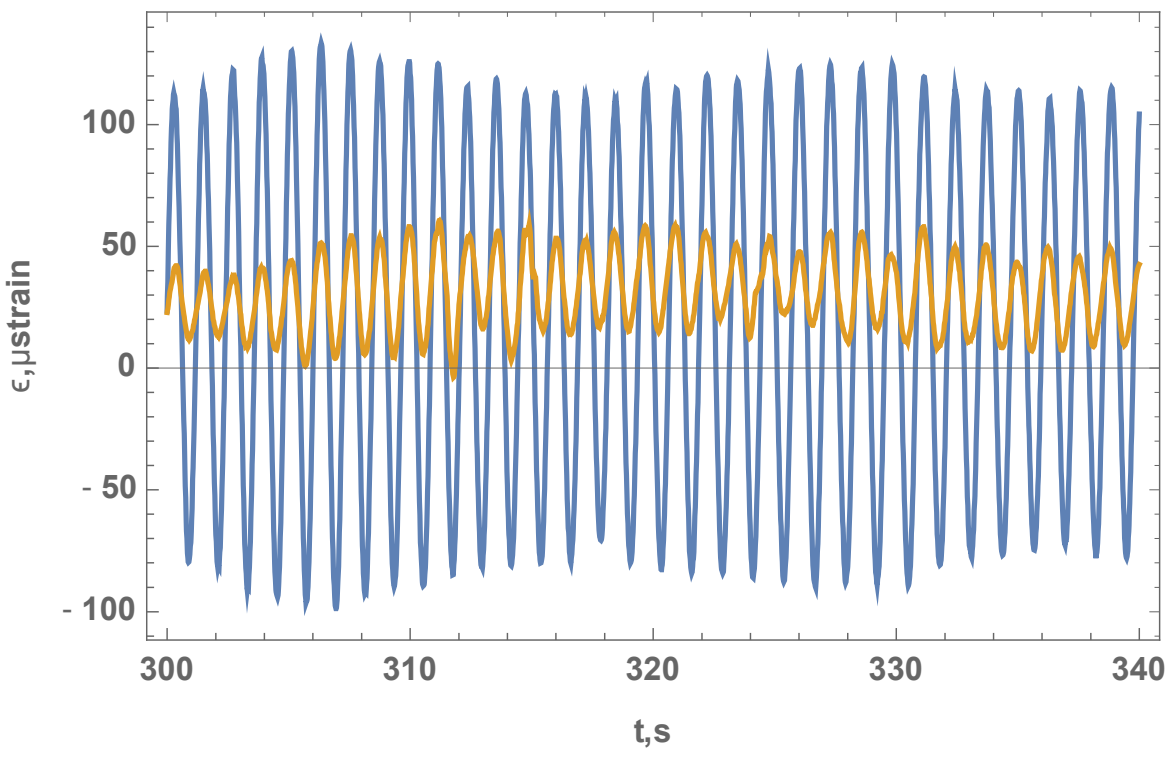

d)

Figure 23. Examples of the records of FBGS1 (blue lines) and FBGS5 (yellow lines) in the tests T16_08 (a,b) and T16_08_mov (c,d).

Figures 24 shows the Fourier spectrums of strains recorded in the locations FBGS1 and FBGS5 (Fig. 1b) in three tests (T16_06, T16_08, and T16_10) with fixed ice (blue lines) and three tests (T16_06_mov, T16_08_mov, and T16_10_mov) with moving ice (yellow lines). Figures 25 shows the Fourier spectrums of strains recorded in the locations FBGS1 and FBGS5 in three tests (T17_06, T17_08, and T17_10) with fixed ice (blue lines) and three tests (T17_06_mov, T17_08_mov, and T17_10_mov) with moving ice (yellow lines). The Fourier spectrums of the Qualisys data recorded in the same tests are shown in Fig. 14. It is obvious that conclusions about the influence of ice motion on the spreading of spectrums around the wave frequency and an increase of wave damping can be obtained based on Fig. 24a and Fig. 25a (see similar results in the discussion of Fig. 14). Figures 24b and 25b confirm that these effects are also seen at a distance of 50 m from the ice edge in the location FBGS5. There is a spectral peak on the blue line corresponding to the second harmonics at the frequency of slightly above 1.2 Hz in Fig. 23a. The yellow line has several peaks around this value in Fig. 24a. Figure 24b shows a spectral peak on the blue line at the same frequency, of around 1.2 Hz, and there are no peaks on the yellow line. Again, this supports a hypothesis of stronger damping of shorter waves in the experiments with moving ice.

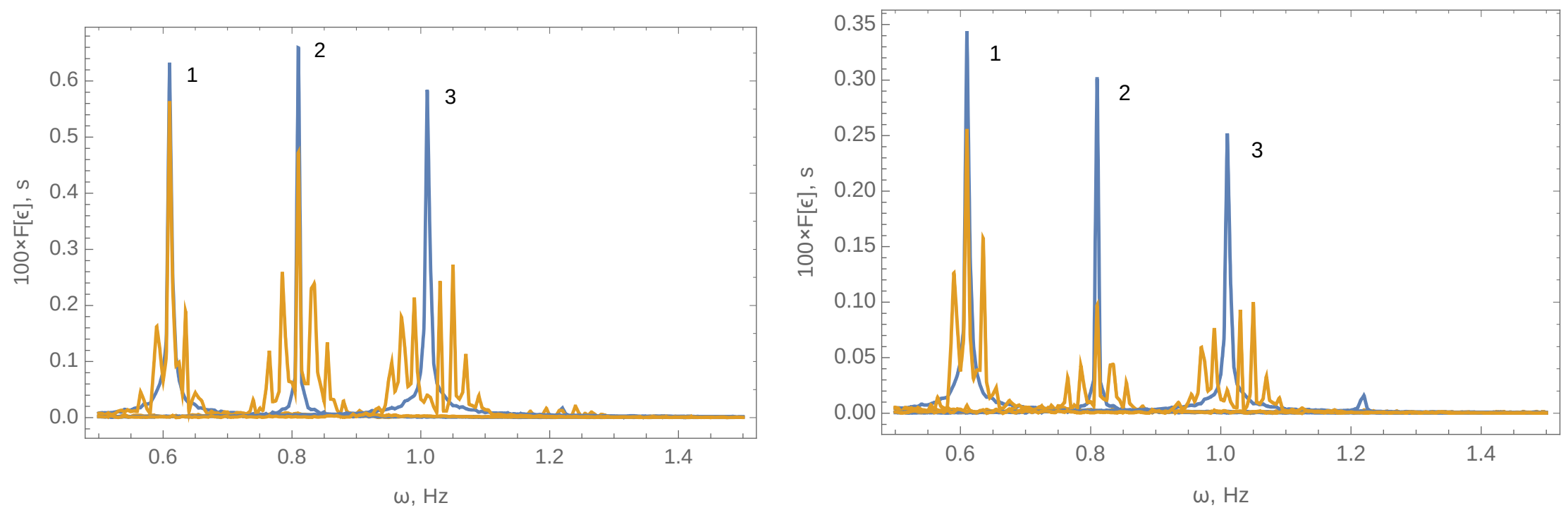

a) b)

Figure 24. Fourier spectrums of the longitudinal strains recorded by the sensors FBGS 1 (a) and FBGS5 (b) in the tests (T16_06, T16_08, and T16_10) with fixed (blue lines) and in the tests (T16_06_mov, T16_08_mov, and T16_10_mov) with moving (yellow lines) ice.

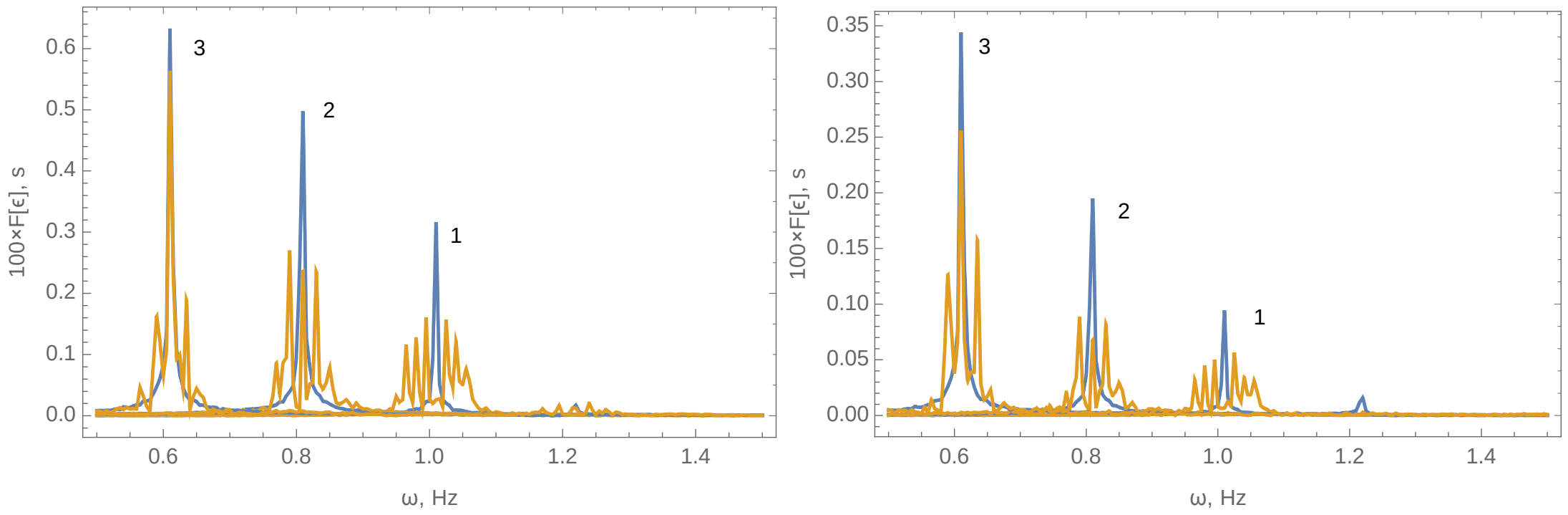

Figure 25. Fourier spectrums of the longitudinal strains recorded by the sensors FBGS 1 (a) and FBGS5 (b) in the tests (T17_06, T17_08, and T17_10) with fixed (blue lines) and in the tests (T17_06_mov, T17_08_mov, and T17_10_mov) with moving (yellow lines) ice.

Strains recorded by FBGS1 and FBGS5 in the tests T17_06, T17_07 and T17_07_01 are shown in Fig. 26 versus the time. Qualisys data recorded in the same tests are shown in Fig. 16. Figure 26a and 26c show results from one set of tests, and Fig. 26b and 26d from another set of tests, showing (as with Fig. 16) that this phenomenon occurs repeatedly. Blue lines in Fig. 26 correspond to the strains recorded before the formation of the non-through crack near the ice edge. The yellow lines correspond to the strains recorded in two tests performed in a row after the crack formation. From Fig. 25 it is evident that the formation of the non-through crack increases wave damping significantly. The effect is seen in both measurement locations, FBGS1 and FBGS5. The black lines in Fig. 26 show that resulting stretching of the ice is greater in the location FBGS5 in comparison with the location FBGS1 (Fig. 1b).

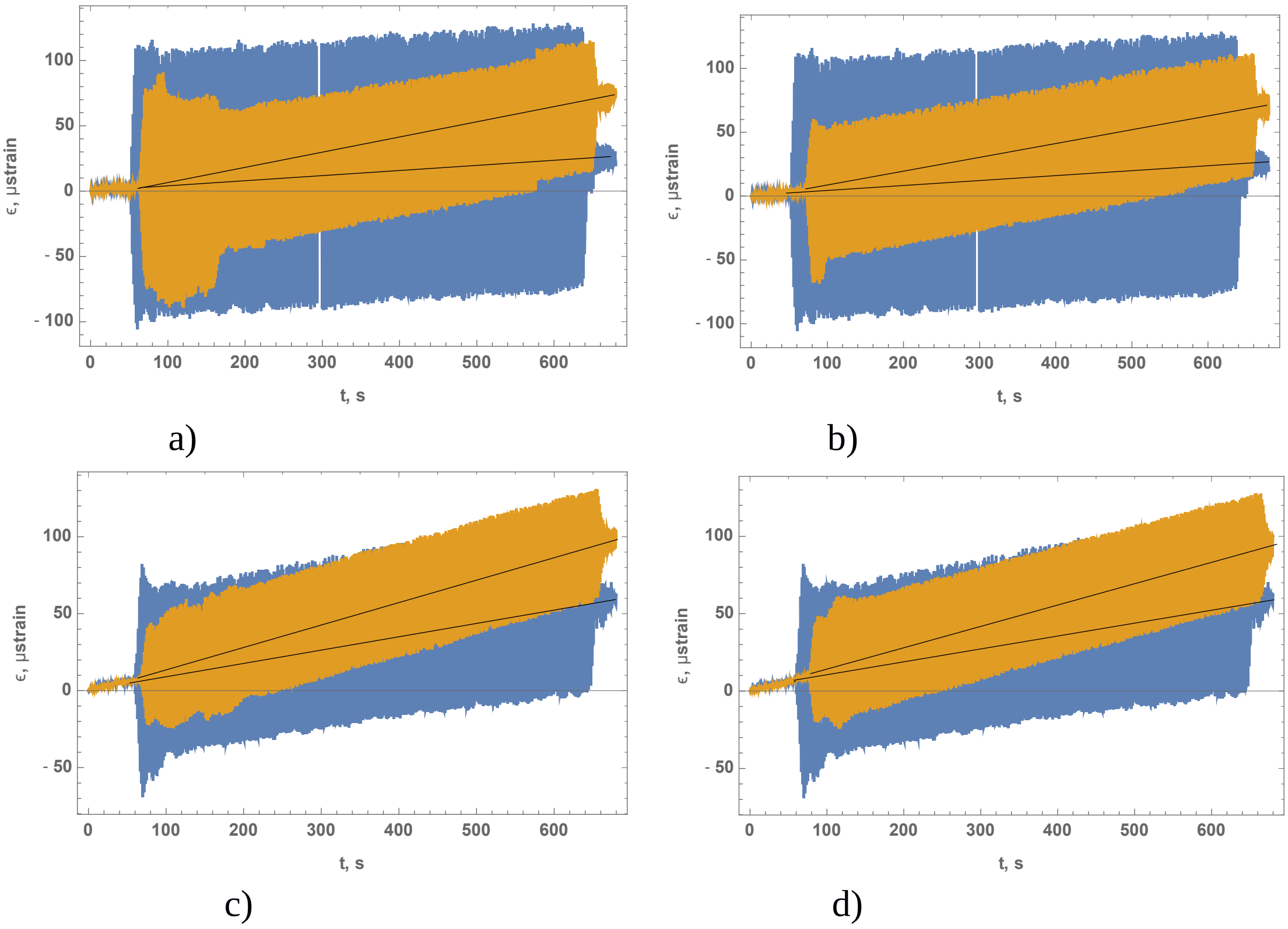

Figure 26. Records of FBGS1 (a,b) and FBGS5 (c,d) in the tests T17_06 (blue lines), T17_07 (a,c; yellow lines) and T17_07_01 (b,d; yellow lines).

FBGT sensors were installed near FBGS sensors at distances of 19 m (location FBGT1 in Fig. 1b) and 50 m (location FBGT2 in Fig. 1b) from the ice edge. FBGT sensors were initially calibrated by placing them into fresh melt water with the temperature of 0°C. Figure 21 shows schematic of FBGT installation and the locations of the temperature sensors on the FBGT. Figure 27 shows records of the air (sensors 1-3), ice (sensors 4-7) and water (sensors 8-11) temperatures in the locations FBGT1 and FBGT2 (Fig. 1b) during the tests T16_08 and T16_08_mov. Figure 23 shows FBGS records from the same tests. Figures 27 a,b correspond to the test with fixed ice. Figures 27 c,d correspond to the test with moving ice. The ice and water temperatures in the test with fixed ice are stable during the test. In the tests with cyclic motion of the ice, we see synchronous changes of the air, ice and water temperatures in the boundary layers. This effect may be related to the water temperature gradients in the tank. Displacement of the ice to a new location causes changes of the water temperature around the FBGT sensors mounted to the ice. These temperature fluctuations penetrate upwards through the holes where the FBGT sensors are installed, due to the under-ice turbulence excited by the cyclic motion of the ice.

Figure 27b shows a decrease of the vertical temperature gradient over the test in the location FBGT1, while Fig. 27c shows a relatively stable vertical temperature gradient in the location FBGT2. This difference can be explained by the influence of waves on the vertical mixing in the boundary layer. The wave amplitude in location FBGT1 is greater than the wave amplitude in location FBGT2. Therefore, the water mixing in the location FBGT1 is stronger than in the location FBGT2. Graphs in Fig. 27c show an increase of the air temperature and the ice temperature measured by the sensors 1-7 towards the end of the test, while the final water temperatures measured by the sensors 8-12 are similar to their initial values. This effect is not visible in Fig. 27d.

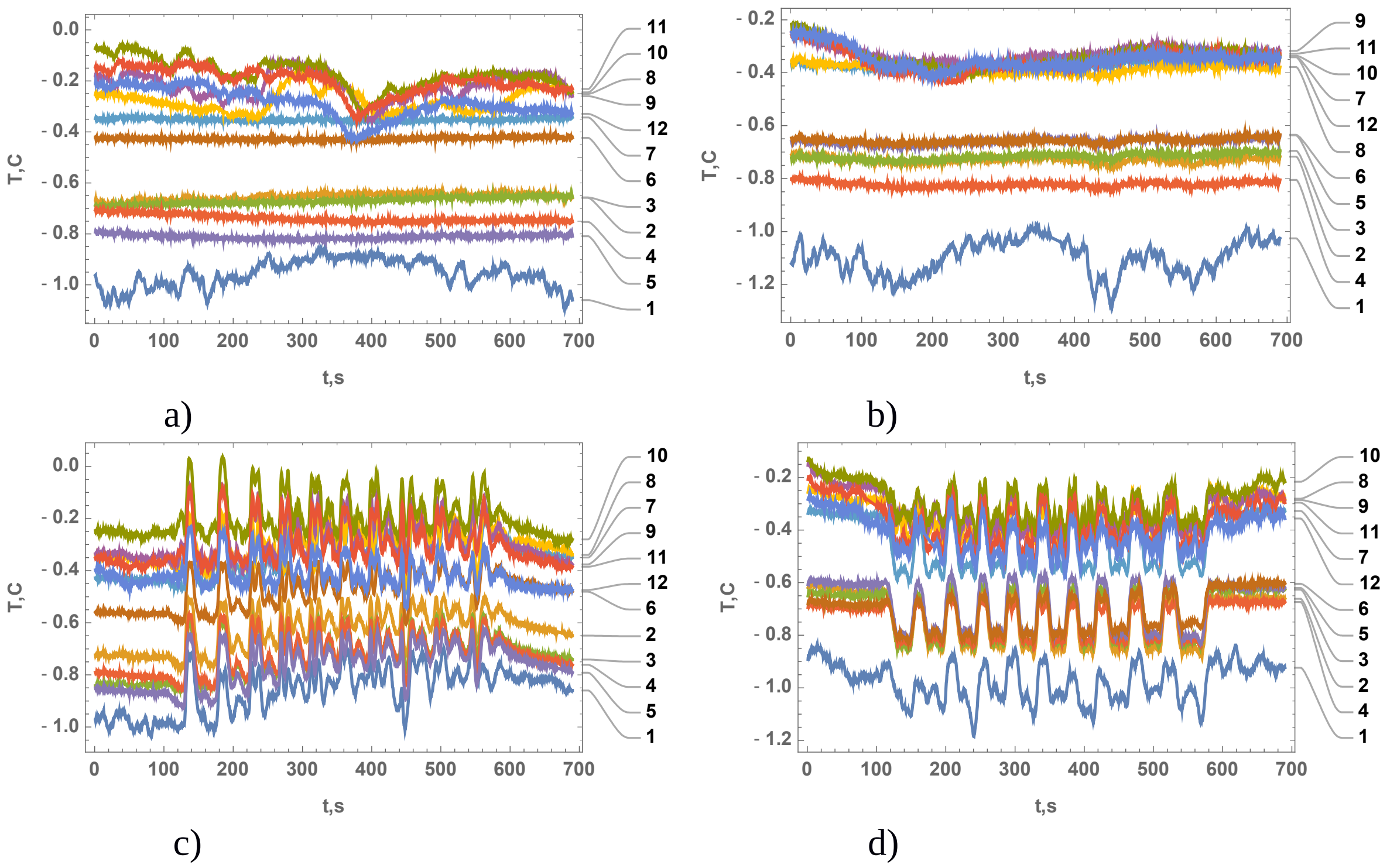

Figure 27. Temperature records in the tests T16_08 (a,b) and T16_08_mov (c,d) in the locations FBGT1 (a,c) and FBGT2 (b,d).

## 8. Deployment and selected results of AE sensors

When ice is under strain, crack formation and dislocation movement make low amplitude sounds. By recording these sounds, it is possible to make inferences about the state of the ice. Here, we present results of acoustic monitoring of the ice sheet as it is deformed by water waves. During all wave-tank experiments, acoustic emissions (AE) were recorded from eight sensors using a Vallen AMSY5 system. We use PZT-5H compressional crystal sensors (Boston Piezo-Optics inc., 15mm diameter, 5mm thickness, 500kHz centre frequency), potted in epoxy, and frozen directly onto the ice surface. Transducers are secured by pipetting 50ml of cold fresh water onto the ice surface, placing the transducer onto the freezing water, and solidifying with a cooling spray. The eight transducers used in this experiment are numbered 1-4 (around 20m from the wavemaker) and 5, 6, 8 and 9 (around 50m from the wavemaker). The signal from these transducers is amplified locally by Vallen preamps (40dB gain), and this amplified signal is then transmitted to a central processing unit. The locations of the transducers are marked by red squares in Fig. 1, and the sensors and their amplifiers are shown in context in Fig. 28. Amplifiers are supported on a movable shelf so that only the piezo crystals are in contact with the ice sheet.

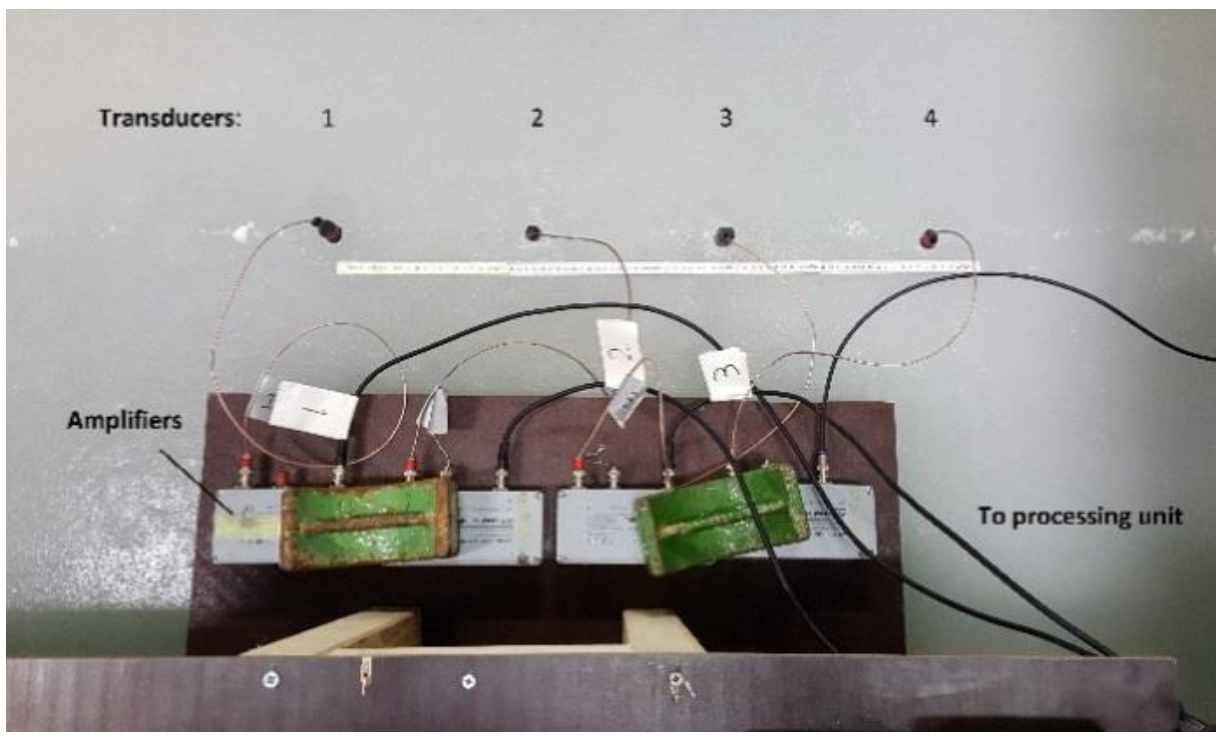

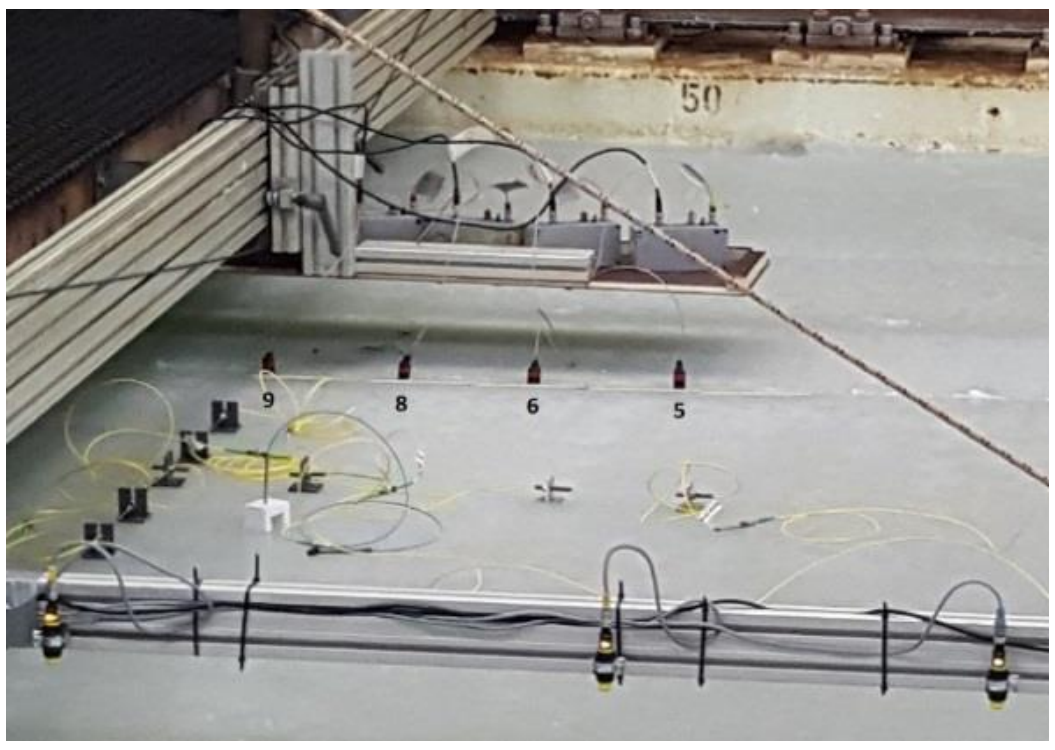

a) b)

Figure 28. (a) Acoustic transducers (1-4) frozen onto the ice, and connected to local preamplifiers, which transmit amplified signals to a central processing unit. (b) AE transducers (5, 6, 8 and 9), FBG sensors on the ice (yellow cables, square brackets) and ice height sensors (mounted on an arm in the foreground.)

The Vallen system records signals from all eight transducers. When the signal reaches a given threshold (40 dB here, corresponding to 0.01mV amplitude) a "hit" is recorded: the system records the time and maximum amplitude of the hit, along with a 400μs windowed transient recorded at 5MHz (i.e. a 400μs recording of the voltage on that channel, beginning 50μs before the threshold was triggered). Typical hit rates are 10-1000 hits per second, depending on the nature of the experiment.

Due to the low amplitude signals being measured, it is difficult to eliminate noise at the hardware level. Therefore, noise is removed in post-processing. To distinguish between signal and noise, data was recorded during a flexural strength test, where the failure in the ice was clear and could be accurately timed. Figure 29 shows, in the top left-hand corner, a signal which corresponds to a single acoustic event within the ice during flexural failure. The figure in the bottom left-hand corner shows a signal which corresponds to noise, recorded several minutes after failure had occurred. Frequency analysis of these signals (and other similar signals) shows that transients due to ice failing tend to have peak frequency components in the range 100-160kHz, while noise signals have peaks at higher frequency. Given this distinction, we set an upper limit of 170kHz on the peak frequency of any data received, and discard hits with higher frequency as presumed noise. It is not feasible to individually check hundreds of thousands of transient signals recorded, but spot checks suggest that the data that is kept (f<170kHz) are qualitatively similar to the signal in the top left of Fig. 29 (a rapid rise, triggering the hit, followed by a slower but clear decay).

An illustrative set of results from the test T16_05 is shown in Fig. 30. The figure shows plots of hit amplitude vs time, recorded on each of the eight channels, for the entire experiment (LHS) and for a 30s window (RHS). Figure 30 shows repeated strong hit data from channels 1-3, and lower amplitude and less frequent hits on channels 5, 6, 8 and 9. This supports a hypothesis that microcracks and non-through cracks develop in the ice during wave loading. Channel 4 recorded fewer hits than all other channels (and more noise), probably because of a faulty transducer. Some preliminary results are worth noting:

- the stronger signal in channels 1-3 is because the wave amplitude, and hence the ice deformation, is higher here. Correspondingly, the amplitudes and numbers of hits recorded on channels 5, 6, 8 and 9 are lower since the waves are significantly damped at this end of the tank.
- Channels 1-3 show a signal which is periodic with the same frequency as the wavemaker. This periodicity is less clear in the signals from the far end of the tank, although further analysis across our recorded data may detect periodicity in the signals recorded by these transducers.
- On each channel there is a strong signal after the wavemaker starts, which decays after the first ~30s. This suggests that there is more acoustic activity when the ice starts to deform, and that this activity decreases with continued deformation caused by wave actions.
- There is notable variation within channels over the duration of the experiment: for example, on channel 1, after an initial period of relatively intense AE (~60-120s), there's a period of less intense emissions, and AE activity then rises again and reaches a peak between 300 and 400s. Patterns on other channels are qualitatively similar but quantitatively different, suggesting that periods of intense AE may represent local cracking close to individual transducers.

Figure 31 shows AE records (hit amplitude as a function of time) from the experiments with fixed (tests T16_08 and T17_08) and moving ice (tests T16_08_mov and T17_08_mov). It is evident that number of hits (and the typical amplitudes of those hits) recorded on channels 1-3 (see Fig. 1b) in the experiments with moving ice is less than in the experiments with fixed ice. This is in keeping with the evidence of the Qualisys data (Fig. 14) and the FBGS sensors (Fig. 24): the experiments with moving ice show a reduction in wave amplitude (and hence in cracking of the ice, and therefore in AE hits). The data recorded by sensors 5-8, in the far end of the tank (Fig. 1b), are too sparse to be shown in the figures. Hit counts for experiments with fixed and moving ice are shown in Table 5. The experiments are grouped into twos, where the top of each pair is a fixed-ice experiment and the bottom of each pair is a moving-ice experiment.

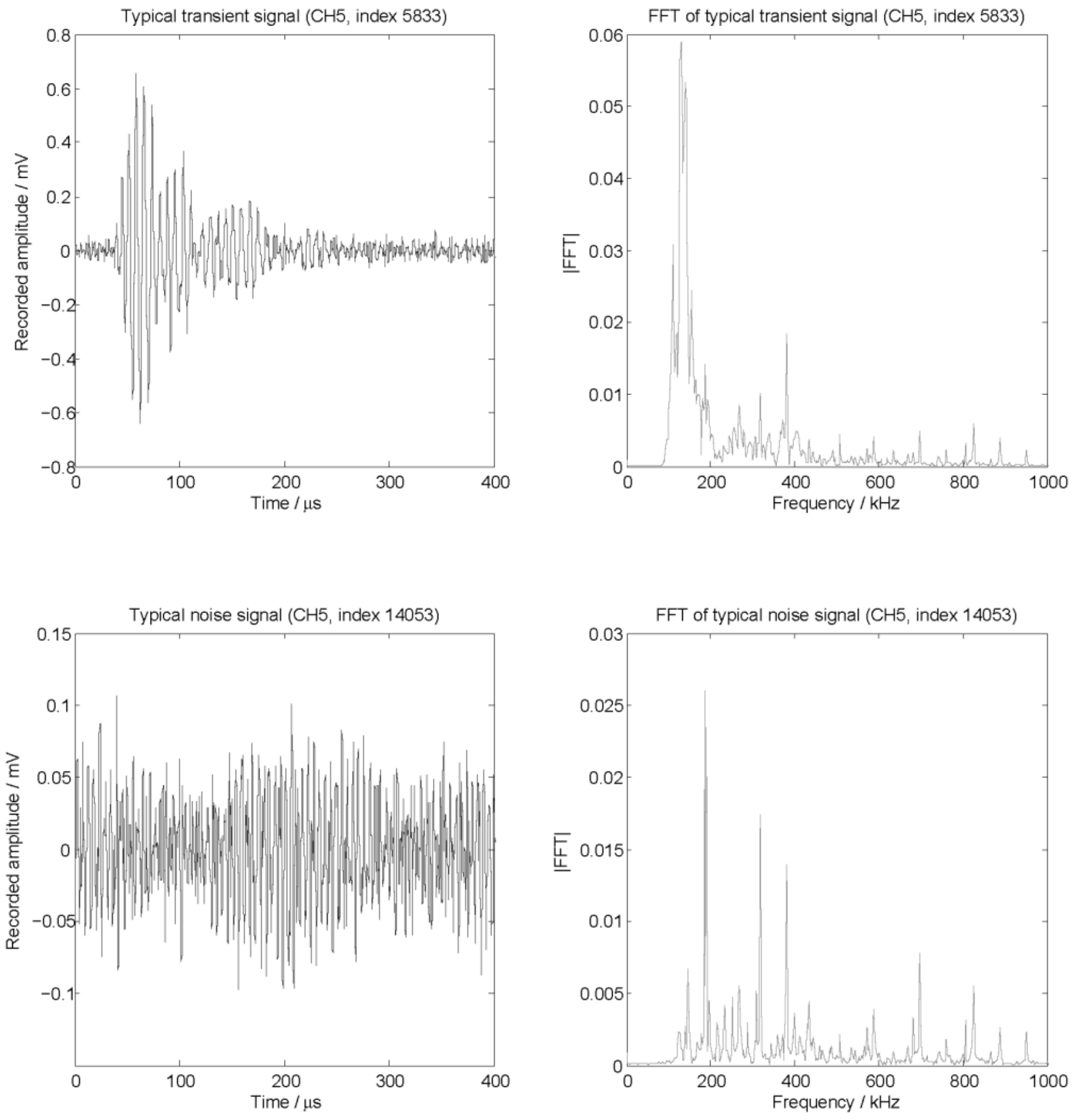

Figure 29. A typical transient signal from a single acoustic event (top left) and a noise signal recorded as a hit (bottom left). FFTs of both signals are shown on the right-hand side.

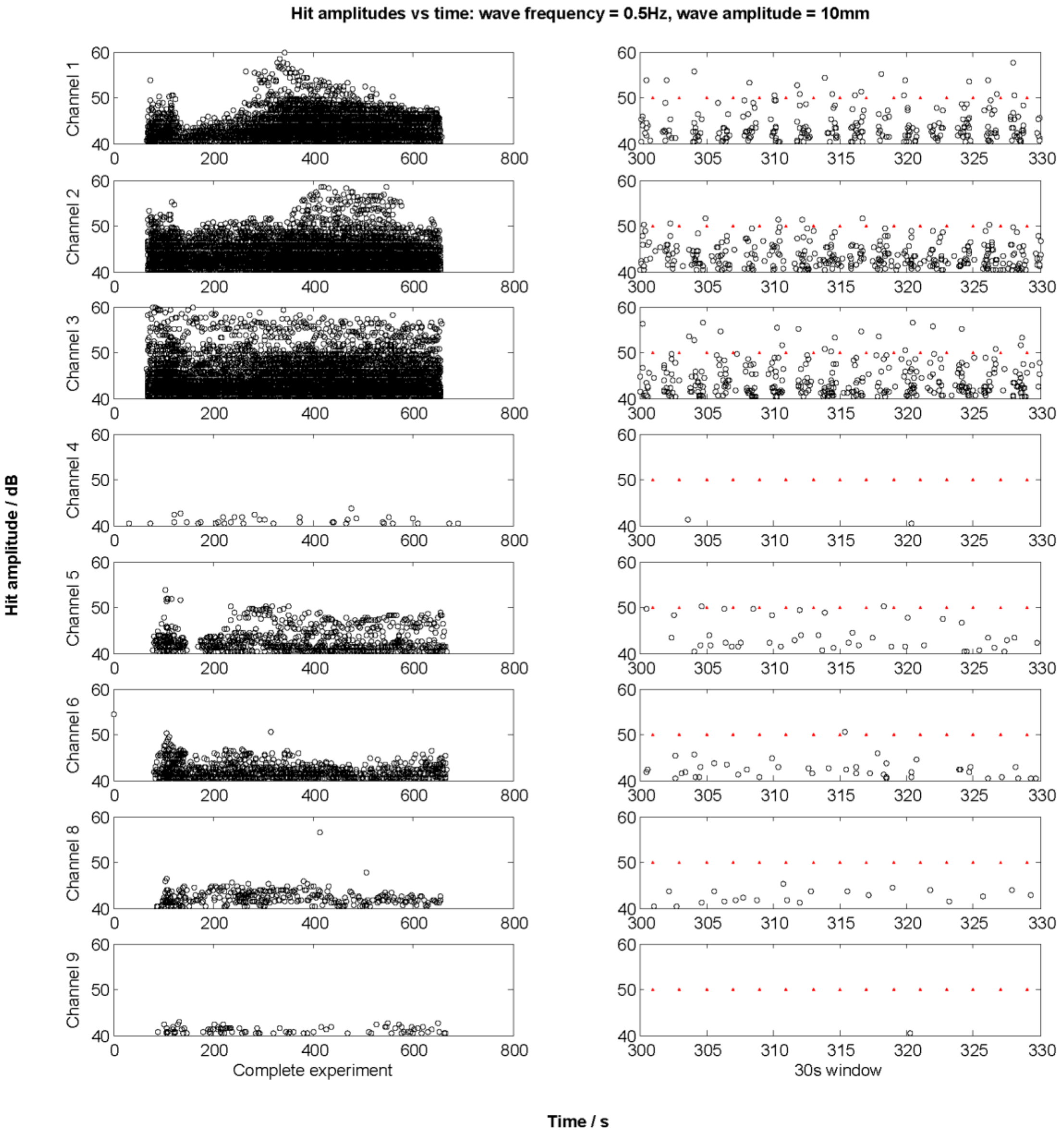

Figure 30. Hit amplitude (on a decibel scale) vs time, for each of eight channels, shown for the entire experiment (left hand side: the wave maker runs from about 60s to 660s, and the start and end of the waves can be clearly seen on several channels) and over a 30s window (right hand side). Red markers are shown at the frequency of the experimental waves (0.5Hz) in the right-hand graphs.

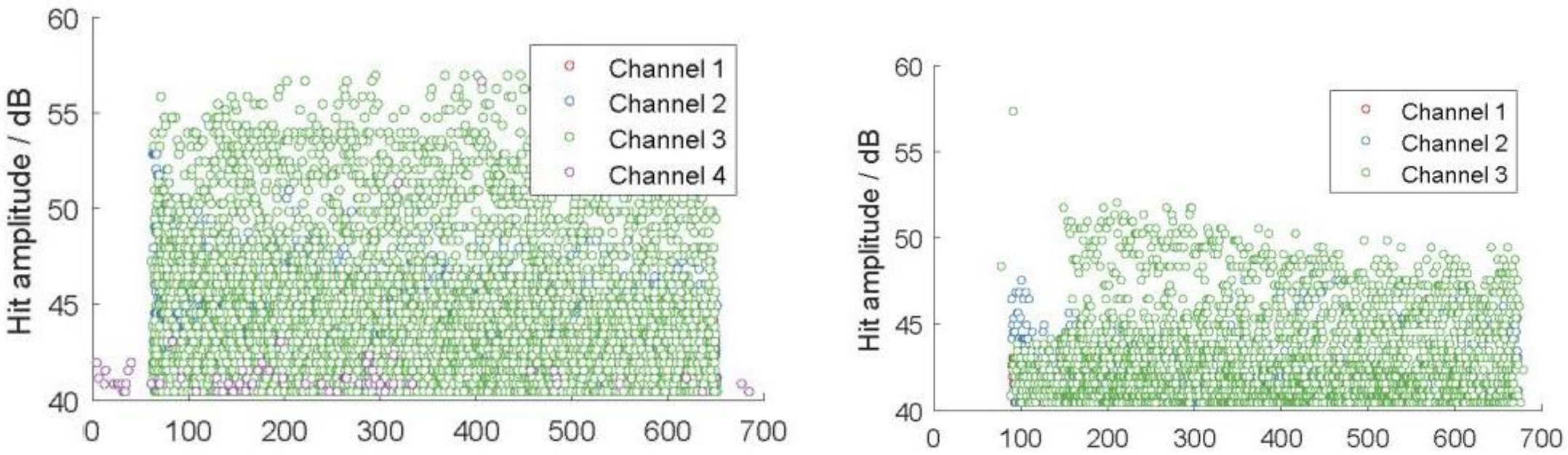

a) b)

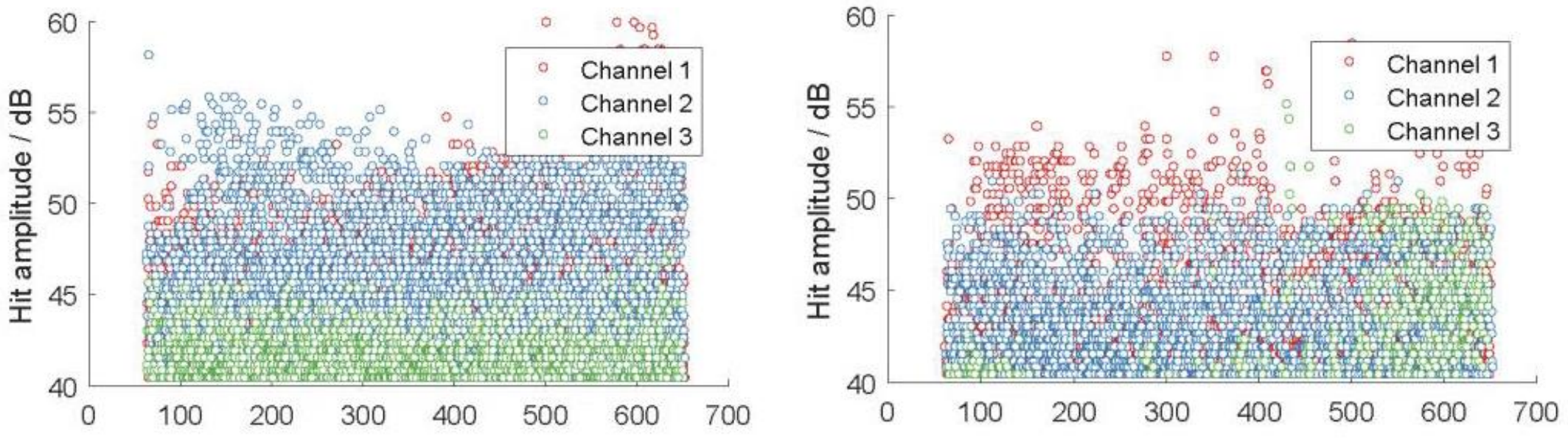

c) d)

Figure 31. Hit amplitudes as a function of time, across three channels, for the tests with fixed ice T16_08 (a) and T17_08 (c) and moving ice T16_08_mov (b) and T17_08_mov (d).

Table 5. Characteristics of AE records in the tests with moving and fixed ice.

| Experiment | Hit Count | | | | | | | |
|---|---|---|---|---|---|---|---|---|
| | Ch1 | Ch2 | Ch3 | Ch4 | Ch5 | Ch6 | Ch7 | Ch8 |
| 16.01; 8Hz;1cm_fix | 689 | 4551 | 6005 | 65 | 3 | 6 | 12 | 0 |
| 16.01; 8Hz;1cm_mov | 212 | 975 | 2540 | 0 | 1 | 1 | 22 | 0 |
| 17.01; 8Hz;1cm_fix | 831 | 6659 | 1301 | 0 | 1 | 0 | 4 | 1 |
| 17.01; 8Hz;1cm_mov | 351 | 2277 | 422 | 0 | 0 | 0 | 5 | 1 |
| 17.01; 6Hz;1cm_fix | 948 | 5166 | 1442 | 0 | 1 | 1 | 6 | 6 |
| 17.01; 6Hz;1cm_mov | 533 | 3184 | 1425 | 0 | 0 | 1 | 5 | 0 |

## 9. Ultrasonic gauges deployment and data analysis

Ultrasonic gauges are useful for measuring wave elevation in both water and ice. Unlike pressure sensors that have been previously used at the HSVA ice wave tank (see, e.g., Wang and Shen, 2010, and Zhao and Shen, 2015), they are not affected by the exponential decay of the wave water pressure field with depth, making careful measurements and calibration easier to perform. Therefore, they are an ideal tool for monitoring both wave propagation and damping in large facilities. In the following paragraphs, we present the experimental ultrasonic gauges setup used in the course of the measurements campaign. All the details of the logging system are released as open source material.

A total of 16 ultrasonic gauges (model S18UUAQ from Banner Engineering) were deployed along the ice wave tank at HSVA. Those gauges have a resolution of 0.5 mm, and are logged at a sampling frequency of 200 Hz. The gauges were deployed as 4 groups of 4, and are numbered from 1 to 16 following the direction of propagation (groups are therefore composed of gauges 1 to 4, 5 to 8, 9 to 12, and 13 to 16). All groups of gauges are located over the center-line of the wave tank, and aligned with its longitudinal axis. Each group was recorded by a separate logger box, located in the vicinity of each group, and connected to a computer. The spacing between consecutive gauges in each group was set to 0.8, 0.5 and 0.7 m, from front to rear gauge. Due to practicalities in the setup, arrays 1 and 2 are aligned rear to from with the direction of wave propagation, while arrays 3 and 4 are aligned front to rear. Non-constant spacing is chosen to alleviate possible aliasing effect if one should attempt to estimate wavelength from gauges measurements. The position of groups 1, 2, and 3 remained the same for both measurement weeks (distance from the paddle of 6.5, 12.5, and 23.5 m, respectively), while group 4 was slightly moved between week 1 (56 m) and 2 (48 m). The position of all sensors, including wave gauges, is indicated in Fig. 1b.

Each logger box is composed of an Arduino Mega board, voltage dividers to convert the 0 to 10 V output of the gauges into a 0 to 5 V signal, and a connection to the trigger signal. Trigger signal is received from the paddle control system, therefore the gauges are synchronized with all other instruments.

Gauge calibration is performed in two steps. First, before the gauges are positioned over the wave tank, each of them is manually calibrated to enforce a sensing range of 5 to 25 cm. The gauges are then set up so that the distance from the mean water level is 15 cm. Finally, a series of wave gauge measurements are taken while varying the water level in the whole wave tank by a fixed amount. This lets us fit a linear calibration curve to the output of the gauges, therefore making sure that effects such as gauge tilt and systematic errors in the logging system are eliminated.

Typical results obtained from the wave gauges, corresponding to a wave amplitude of 1.5 cm and wave frequency of 0.7 Hz, are presented in Fig. 32. Calibrated wave elevation data measured by gauges 1, 5, 9 and 13 (i.e., the first gauge in each group) is presented alongside an illustration of the wave exponential attenuation. As visible in the calibrated data, wave elevation measurements are of good quality and individual waves, as well as damping of the wave amplitude along the wave tank, are clearly visible. Elevation data can be further processed to obtain the wave elevation by integrating the Fourier spectrum of the waves around the peak frequency of the waves. In Fig. 32b we compute the Fourier spectrum based on 5 minutes of wave data corresponding to the middle of the time series using a Hamming window, and we integrate on an interval of 0.10 Hz around the wave peak frequency. This methodology is similar to the one presented in Sutherland et al. (2017). Exponential wave attenuation is clearly visible in the ice covered region.

In addition to the expected wave propagation and wave damping, both reflections and 3D effects are present in the wave tank. This is visible through the variations in wave amplitude obtained from different gauges in the same group (see Fig. 32b), and also directly in the calibrated elevation data (and then corresponds to a modulation of the wave amplitude with time).

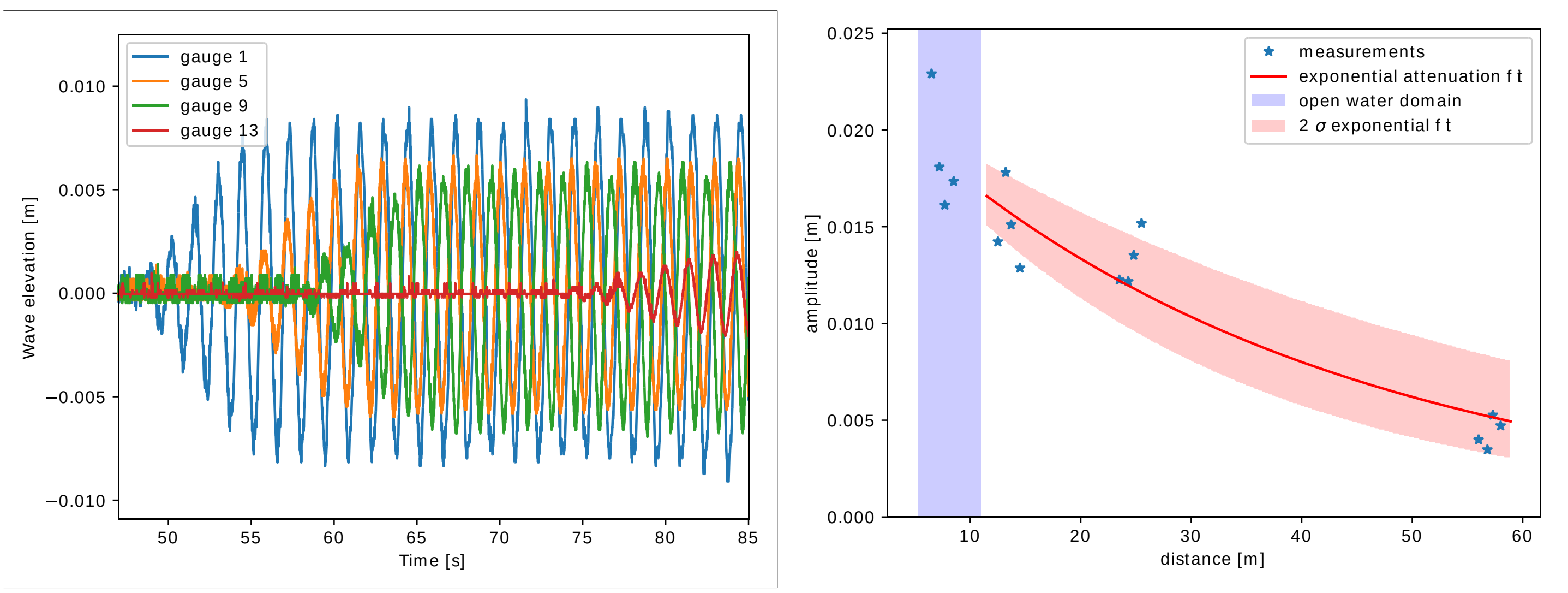

Figure 32. Example of calibrated data obtained from gauges 1, 5, 9 and 13, corresponding to incoming waves of amplitude 1.5 cm and frequency 0.7 Hz (a). Corresponding wave attenuation and exponential fit (b). The wave amplitude is computed by integrating the Fourier spectrum of the waves around the wave peak frequency.

The effect of ice motion, when present, is clearly visible as a modulation in the wave signal as visible in Fig. 33a. Similar effects of wave modulation in the moving ice were registered by Qualisys system and FBGS sensors (Fig. 23 c,d). However, it is challenging to know what the effect on damping is due to the 3D effects, and the fact that the properties of the ice and ice edge change between the runs due to the effect of the waves. Despite these difficulties, the effect of ice motion on the wave amplitude some distance inside the ice is clearly visible. Fig. 33b shows both the wave signal, detrended from the mean offset variations induced by the ice motion using low-pass filtering, and its Hilbert transform, for the first gauge of the third array (gauge 12), in the case with fixed ice (T16_06) and moving ice with the same incoming wave conditions (T16_06_mov). As it is visible in Fig. 33b from the results of the Hilbert transform, the ice motion periodically induces increased damping. This can be a sign that the interaction between the ice motion and the boundary layer under the ice may create additional damping.

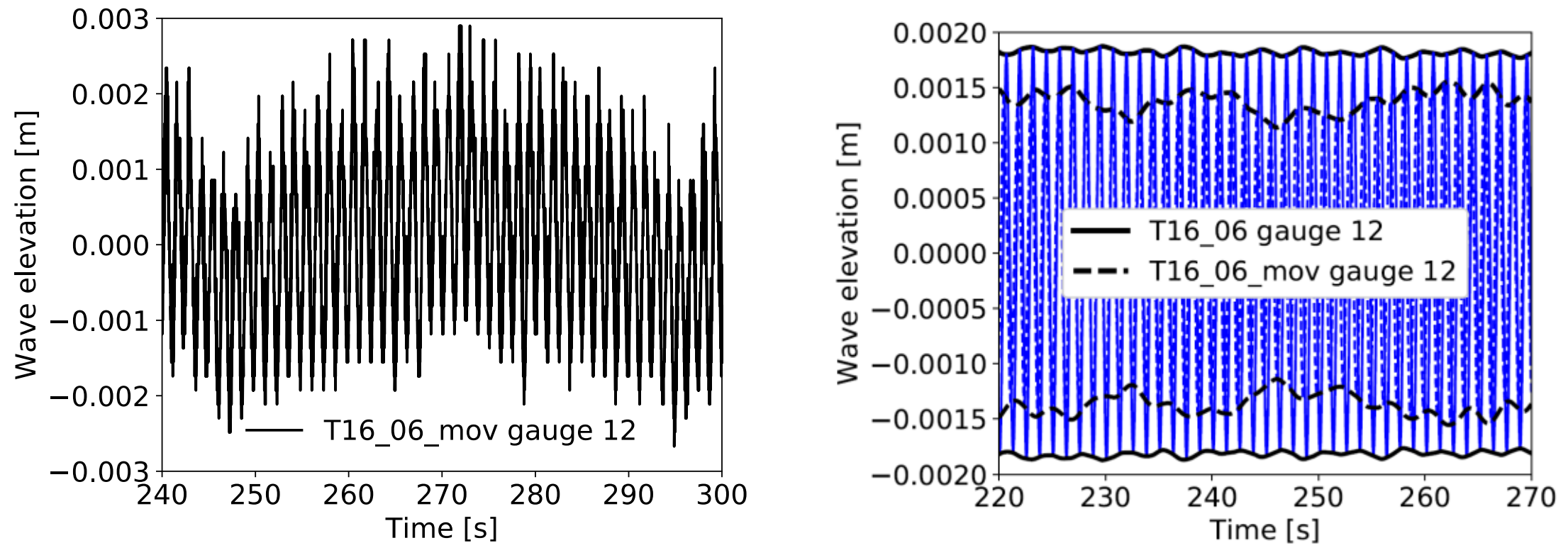

a) b)

Figure 33. Wave modulation registered by US sensors in the test T16_06_mov with moving ice (a) and effect of wave modulation on the wave amplitude below the ice in the test T16_06_mov in comparison with the test T16_06 with fixed ice (b).

## 10. Water pressure measurements

Eight water pressure (WP) sensors were installed along the tank, in locations WP1 (one sensor), WP2 (one sensor), WP3-5 (three sensors), and WP6-8 (three sensors) (Fig. 1). All water pressure sensors were installed at a water depth of 15 cm. Water pressure was measured at a rate of 200 Hz, with each sensor logged into an individual channel. All channels were measured with the same data acquisition system and share the same time channel, thus are synchronized. Initial observations of the results show that the difference between the water pressure measurements that were taken relatively close to each other is not significant (at least for basic analysis), thus in the following only one sensor from each location is presented. It is always the first in the row of three (e.g. WP 3 and WP 6). In Fig. 34 the complete time series of water pressure are shown for the experiments with fixed ice (a) and cyclically moving ice (b). The motion of the ice sheet is clearly visible in the time series (b) as a modulation on the signal. Figure 35 shows a closer look at the time series from the same experiments. The waves are clearly visible in all of the time series (in open water and under the ice). The first impression is that in both cases (fixed and moving ice sheet) the ice sheet has an impact on the amplitude of the water pressure measurement but not the frequency.

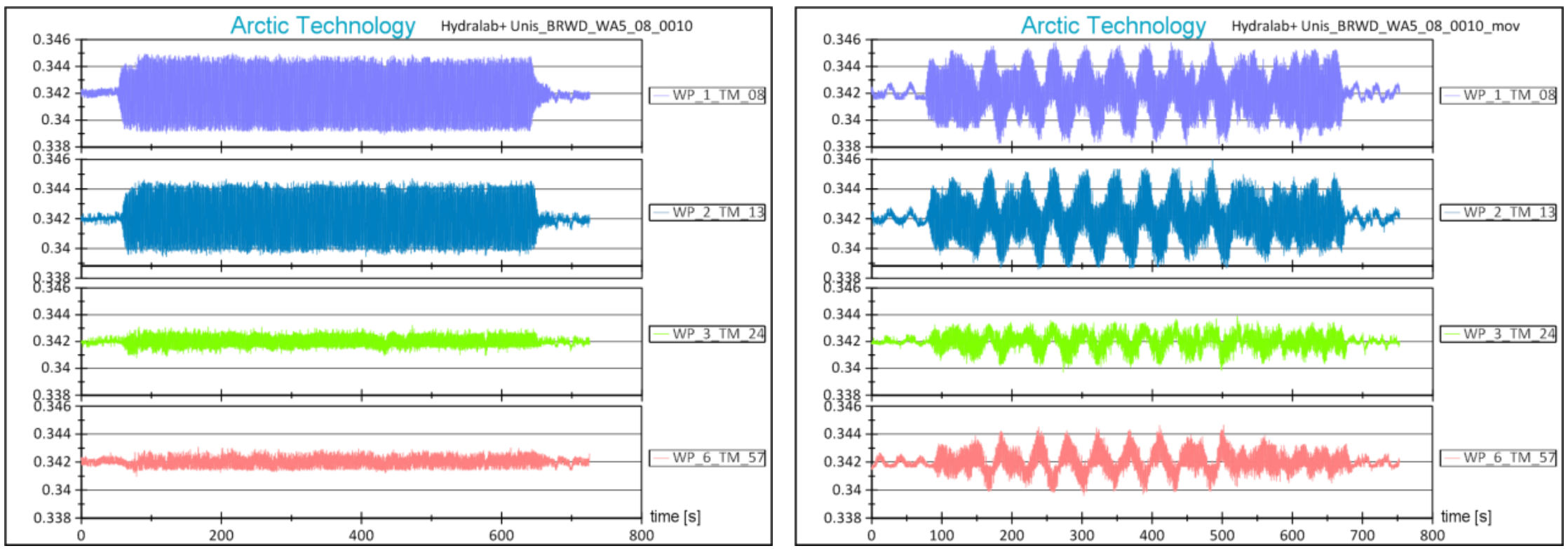

a) b)

Figure 34. Time series of WP1 (light blue), WP2 (dark blue), WP3 (green), and WP6 (pink) for the test T16_08 with fixed ice (a) and in the test T16_08_mov with moving ice (b).

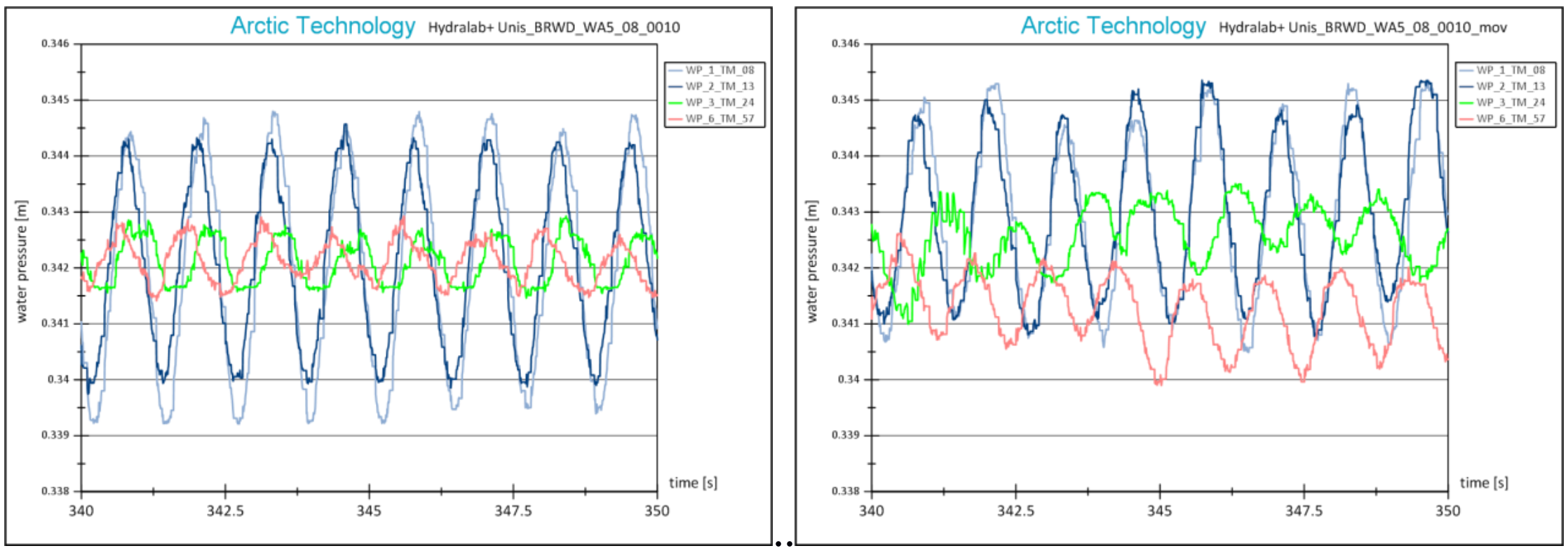

a) b)

Figure 35. Time series of the records of WP1 (light blue), WP2 (dark blue), WP3 (green) and WP6 (pink) in the test T16_08 with fixed ice (a) and in the test T16_08_mov with moving ice (b).

The wave damping can be analyzed using Fourier spectrums shown in Fig. 36. Figures 36a,b correspond to three tests (T16_06, T16_08, T16_10) with fixed ice and three tests (T16_06_mov, T16_08_mov, T16_10_mov). Figures 36c,d correspond to three tests (T17_06, T17_08, T17_10) with fixed ice and three tests (T17_06_mov, T17_08_mov, T17_10_mov). Numbers 1,2 and 3 show the order of the tests with moving ice similarly to Fig. 24,25. Lines 1 and 2 in Fig. 36a,b show stronger wave damping in the experiments with fixed ice, while Line 3 in Fig. 36a,b and Lines 1,2,3 in Fig. 36c,d show stronger wave damping in the experiments with moving ice. Fig. 36 shows a small spectral local maximum at the frequency of 1.2 Hz in the experiments with main wave frequency of 0.6 Hz.

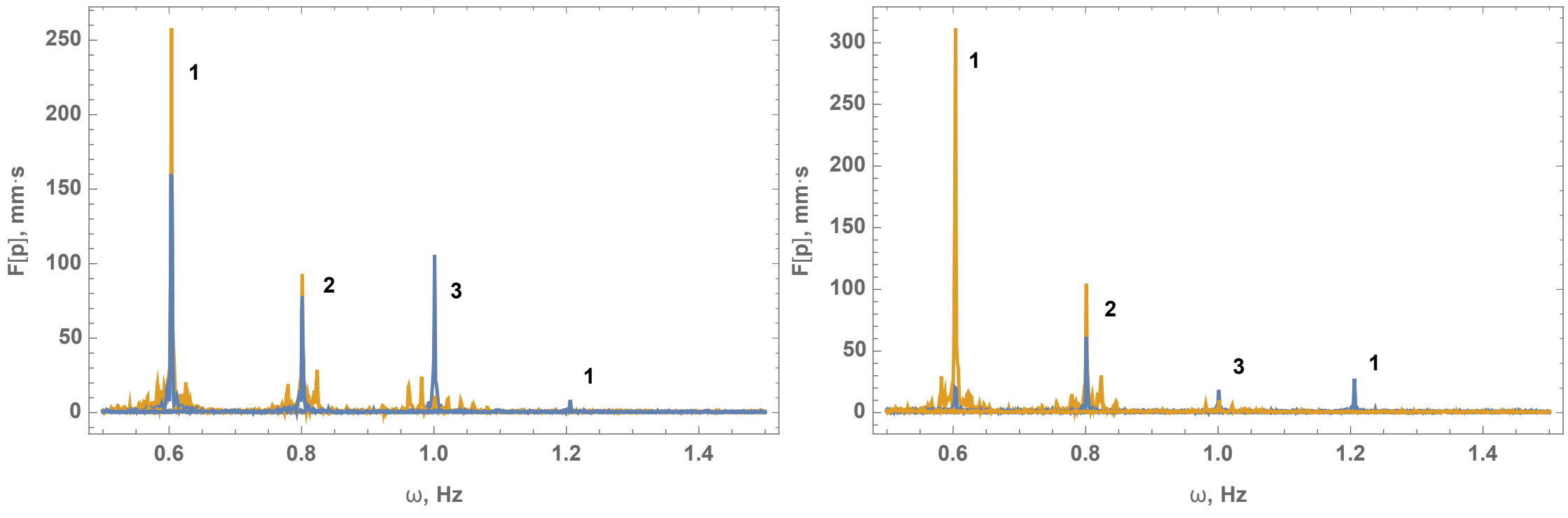

a) b)

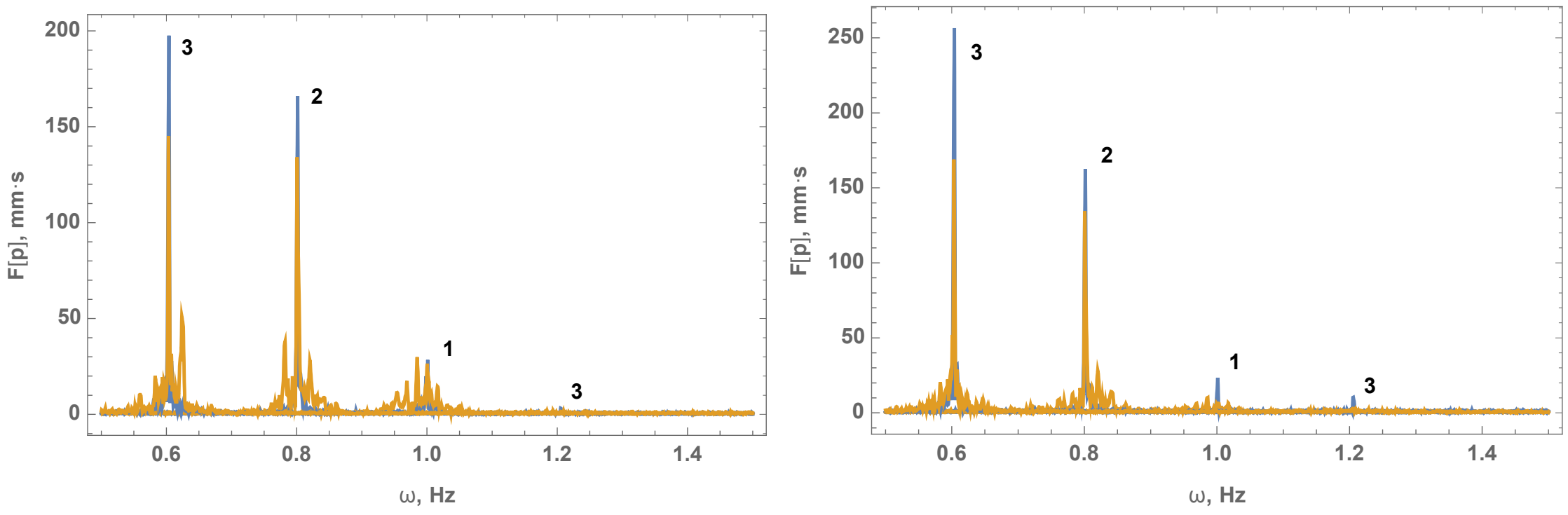

c) d)

Figure 36. Fourier spectrums of the pressure records by WP3 (a,c) and WP6 (b,d) from the tests (a,b) T16_06, T16_08, T16_10 (blue lines), T16_06_mov, T16_08_mov, T16_10_mov (yellow lines) and (c,d) T17_06, T17_08, T17_10 (blue lines), T17_06_mov, T17_08_mov, T17_10_mov (yellow lines).

Figure 37 shows records of the pressure fluctuations by WP1 (blue lines) and WP3 (yellow lines) in the tests T17_06 performed before the formation of the non-through cracks, and two tests T17_07 and T17_07_01 performed after the formation of the non-through crack. Figures 16 and 26 show Qualisys data and FBGS records obtained in the same tests. The blue lines in Fig. 37 show the same wave height of about 6 mm on the open water in all tests, which is smaller than wave height of 1 cm declared by wave maker settings. The yellow lines show that wave amplitudes registered by WP3 in the tests performed after the non-through crack formation are smaller than wave amplitudes recorded by the same sensor WP3 in the tests before the crack formation. This corresponds to the results obtained from the analysis of the Qualisys and FBGS records. Comparison of the blue lines with the yellow lines shows a decrease of wave amplitudes below the ice in comparison with the wave amplitudes on the open water.

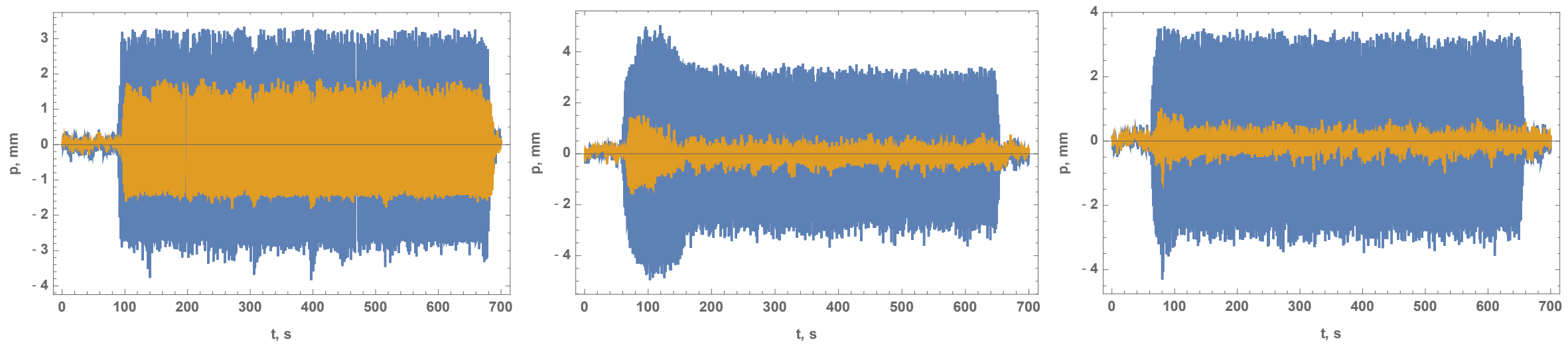

a) b) c)

Figure 37. Records of WP1 (blue lines) and WP3 (yellow lines) during the tests T17_06 (a), T17_07 (b) and T17_07_01 (c) .

## 11. Conclusions

Experimental investigations of flexural-gravity waves, performed in HSVA ice tank, were focused on the study of the physical mechanisms of wave damping. Similarity criteria based on the Froude number and number $a_{fg}$ (characterizing the influence of elasticity on wave dispersion in the experiment) show that our experimental results correspond to waves with periods of about 10 s and smaller, propagating below ice with thickness of 1m and greater in natural conditions. The advantages of using relatively short model waves in the experiment consist of (a) the reduced influence of the tank bottom on the wave characteristics (deep water conditions) and (b) the reduced amplitudes of the waves reflected from the end of the tank. The reduction in reflections occurs because of stronger damping of the waves as they propagate below the ice to the end of the tank. Because the influence of the tank geometry on the waves was small, this gave us the possibility to investigate how the physical effects of wave-ice interaction influence wave damping.

Several physical effects causing wave damping below the solid ice were observed in the experiments including perforation of the ice edge, formation of non-through cracks, ice break up by waves and under-ice turbulence generated by cyclic motion of the ice sheet along the ice tank. Perforation of the ice edge and ice failure near the edge caused 3D effects within the wave-ice interaction. Non-through cracks, produced in the ice by waves, caused strong damping due to the cyclic pumping of the brine up and down through the ice. Wave damping in the region where the ice was broken by the waves and in the region with a net of non-through cracks prevented ice failure in the end of the tank in all experiments. Wave action on broken ice caused floe-floe interactions leading to the smoothing of floe edges and production of slush between floes. Floe-floe interactions could be observed through relative displacements of floes, floe collisions and floe rotation. Observed effects of floe-floe interactions seem much stronger in the tank, where the ice is confined, in comparison with similar effects in the marginal ice zone in the open sea.

Wave action on the ice and water was measured with a number of sensors. Vertical and horizontal displacements of the ice were measured by the optical system Qualisys, by ultrasonic sensors and by water pressure sensors; horizontal in-plane deformations of ice were measured with fiber optic strain sensors; acoustic emission due to microcracking in the ice was measured by compressional crystal sensors; water velocities over the water column and in the under-ice boundary layer were recorded by an acoustic doppler velocimeter and current profiler; and the air, ice and water temperatures were recorded with fiber optic temperature strings. Small wave amplitudes and clean water led to a somewhat low quality of records from the ADV, ADCP and water pressure sensors. The other sensors provided records of good quality, which demonstrated the influence of ice motion and non-through cracks on wave damping, through reduced amplitudes of the ice surface elevation and FBG strains, and reduced numbers and amplitudes of acoustic hits.

The influence of under-ice turbulence caused by the ice drift on the damping of swell has been investigated for the conditions of the MIZ in the Barents Sea, where the eddy viscosity of the ice adjacent boundary layer of the water was calculated using ADV records (Marchenko et al., 2015; Marchenko and Cole, 2017). These field measurements were performed from drift ice in several regions of the Barents Sea. It was shown that the eddy viscosity increases with the increase of mean water velocity relative to the ice. The eddy viscosity was used instead of molecular viscosity in a solution describing the oscillating boundary layer below the ice, induced by waves. In the present study, insufficient quality of the ADV data means it has not been possible to calculate the eddy viscosity. Nevertheless, the ADV and ADCP records, together with FBGT temperature measurements, show higher fluctuations of the water velocities and temperature in the water layer below the ice in the experiments with moving ice. At the same time, results from the Qualisys system, FBGS and US sensors show stronger wave damping in experiments with moving ice. Thus, the experimental results confirm the importance of ice drift and under-ice turbulence for wave damping.

Results from the Qualisys system, FBGS and water pressure sensors show the existence of the multiple harmonics in the records performed during the tests with fixed ice. These multiple harmonics are not seeb in the spectrums of the data from the tests with moving ice. This suggests that the influence of ice drift on wave damping increases with an increase of the wave frequency.

The influence of non-through cracks on wave damping may explain the effect of strong wave damping by solid ice, not yet split into floes, measured in the Barents Sea (Collins et al., 2015). Strong damping occurred when non-through cracks produced brine pumping. The damping decreases immediately when the crack passes through the ice and splits it into individual floes. Non-through cracks may therefore influence wave damping in pack ice. Note also that non- through cracks are formed in pack ice under the influence of thermal expansion (see, e.g., Lewis, 1993).

Strains, measured with FBGS sensors, indicate stretching of the ice sheet during each test. The mechanical influence of bending ice on the brackets where FBG sensors were mounted (Fig. 19) may cause creep, which would be visible as a compression. Stretching may be a result of ice creep initiated by periodical bending. This stretching may cause an extension of non-through cracks leading to the splitting of the ice into individual floes.

An FBG thermistor string measured oscillations of the ice temperature and water temperature below the ice in the tests with moving ice (Fig. 25c,d). Temperature oscillations are not observed in the experiments with fixed ice (Fig. 25a,b). The period of temperature oscillations in the tests with moving ice coincides with the period of ice motion along the tank. The amplitude of the temperature oscillations, about 0.1oC, is similar to the temperature changes over the ice thickness. The ice temperature oscillations can be explained by vertical migration of the water through the ice around the thermistor string. Under-ice turbulence initiated by the ice motion increases water mixing in the boundary layer and helps water to penetrate into the gap between the FBG thermistor string and ice. A small amount of heating of the ice by waves was measured in the tests with moving ice. This suggests that waves influence the characteristics of the boundary layer.

Acoustic emission was recorded in all tests of the experiments. AE amplitudes show the same periodicity as the wavemaker. This suggests that the waves lead to periodic increases and decreases in microcracking within the ice. These microcracks (or the extension of existing cracks) are then recorded as AE hits. Damping along the length of the tank leads to significant reduction in AE hit counts on sensors further from the wavemaker. Acoustic emissions are at a higher level at the very start of experiments, suggesting that healing may occur while the wavemaker is off, and then cracks reopen quickly once the waves restart. Acoustic emissions are significantly reduced in experiments with moving ice, since damping is greater and the ice is deformed less in these experiments.

The data from these tests will be used for future investigation of the rheology of model ice subject to waves. The data will be used to calculate the dependency of wave attenuation rates on the wave ice characteristics, and to identify 3D effects observed during wave propagation below the ice. Results of the experiments performed with broken ice will be used to estimate similar effects in broken ice and to compare wave damping in solid and broken ice. Measurements performed with the Qualisys system and FBGS sensors will be used to specify in-plane deformations in ice with and without cracks. Acoustic emissions data will be used to quantify healing and damage of ice under repeated loading.

Acknowledgements
The work described in this publication was supported by the European Community's Horizon 2020 Research and Innovation Programme through the grant to HYDRALAB-Plus, Contract no. 654110. The authors also wish to acknowledge the support of the Research Council of Norway through the PETROMAKS2 project Dynamics of floating ice and IntPart project Arctic Offshore and Coastal Engineering in Changing Climate. Authors of the paper thank HSVA staff for the help during the experiment, and Hayley Shen for the discussion of project proposals and experimental results.